# A systematic literature review on source code similarity measurement and clone detection: techniques, applications, and challenges

Morteza Zakeri-Nasrabadi[a,1], Saeed Parsa[a,1], Mohammad Ramezani[a], Chanchal Roy[b], and Masoud Ekhtiarzadeh[a]

[a] School of Computer Engineering, Iran University of Science and Technology, Tehran, Iran.

[b] Department of Computer Science, University of Saskatchewan, Saskatoon, Canada.

e-mails: (*morteza_zakeri@comp.iust.ac.ir, parsa@iust.ac.ir, ramezani_mohammad@noor.iust.ac.ir,*

*chanchal.roy@usask.ca,* and *m_ekhtiarzadeh@noor.iust.ac.ir*)

**Abstract**

Measuring and evaluating source code similarity is a fundamental software engineering activity that embraces a broad range of applications, including but not limited to code recommendation, duplicate code, plagiarism, malware, and smell detection. This paper proposes a systematic literature review and meta-analysis on code similarity measurement and evaluation techniques to shed light on the existing approaches and their characteristics in different applications. We initially found over 10000 articles by querying four digital libraries and ended up with 136 primary studies in the field. The studies were classified according to their methodology, programming languages, datasets, tools, and applications. A deep investigation reveals 80 software tools, working with eight different techniques on five application domains. Nearly 49% of the tools work on Java programs and 37% support C and C++, while there is no support for many programming languages. A noteworthy point was the existence of 12 datasets related to source code similarity measurement and duplicate codes, of which only eight datasets were publicly accessible. The lack of reliable datasets, empirical evaluations, hybrid methods, and focuses on multi-paradigm languages are the main challenges in the field. Emerging applications of code similarity measurement concentrate on the development phase in addition to the maintenance.

**Keywords**: Source code similarity, code clone, plagiarism detection, code recommendation, systematic literature review.

## 1   Introduction

Source code similarity measurement is a fundamental activity for solving many code-related tasks in software engineering, including clone and reuse identification [1]–[9], plagiarism detection [10]–[16], malware and vulnerability analysis [17]–[22], and code recommendation [23]–[26]. Almost all machine learning-based approaches used to measure program quality [27]–[30], detect code smells [31]–[34], suggest refactorings [35], [36], fix programs [37], [38], and summarize source codes [24], [39] work with a kind of source code similarity indicator. Hence, it is essential to have a comprehensive view of code similarity measurement techniques, their essence, and their applications in software engineering.

In literature, code similarity is also referred to as code clone or duplicate code [1], [40]–[47]. While code similarity is a broader concept than code clones, cloned codes are one of the main factors affecting software maintenance. Code clones in software systems often result from the practice of copy-paste programming [41], [48], [49]. This process helps programmers quickly reuse the part of code and speed up the development process, while it may lead to large fragments of code clones. Other actions that lead to code clones are language limitations, coding styles, APIs to execute the same rules, and coincidentally implementing the same logic by different developers [6], [48].

There are difficulties and benefits regarding the make or use of code clones in the software development lifecycle. One major challenge with cloned codes is that if a fault is detected in a piece of code, all cloned parts must be identified to check for the possibility of the defect. Rahman and Roy [50] have shown that code clones decrease the stability of programs. On the other hand, several studies have claimed that refactoring duplicate codes are not always desirable. For instance, removing code clones increases dependencies between different components, which may be tightly coupled due to referring to one piece of code instead of separated clones. Hence, many researchers state that clone codes should be at least identified even if no action is performed [51]–[55]. Unfortunately, detecting code clone instances is not straightforward. However, clone types definition and taxonomies sometimes are not clear, and the border between code clone and code similarity is not defined well in the research.

---

[1] Corresponding author. Address: School of Computer Engineering, Iran University of Science and Technology, Hengam St., Resalat Sq., Tehran, Iran, Postal Code: 16846-13114.

Despite the maturity of studies on code similarity measurement, we found no comprehensive survey in the field covering state-of-the-art approaches. This paper proposes a systematic literature review (SLR) and meta-analysis on code similarity measurement and evaluation approaches to classify the technical aspects existing in the field and disclose the available challenges and opportunities. For this aim, we searched the most important digital libraries, resulting in an initial set of more than 10000 articles. After applying the relevant exclusion and inclusion criteria, snowballing, and quality assessment, 136 articles were selected as the primary studies in the code similarity measurement field. We reviewed all primary studies in detail, analyzed and categorized our findings, and reported the results. The main objectives and contributions of our SLR are as follows:

1. We indicate the state-of-the-art techniques, software tools, and datasets proposed for code similarity measurement and clone detection.
2. We identify and classify essential aspects of the available code similarity measurement studies and compare their advantage and disadvantage in different application domains, including clone detection, malware detection, plagiarism detection, and code recommendation.
3. We designate and discuss the existing challenges and opportunities in the current code similarity measurement and the new applications emerging in the field.

Our study indicates various techniques, tools, datasets, and applications related to code similarity measurements. Most importantly, 80 different software source code similarity measurement and clone detection tools are proposed by the researchers, which work based on at least eight different techniques. Most proposed techniques measure the similarity between tokens in the code fragments, while more advanced techniques based on machine learning and hybrid methods are emerging. The lack of public tools and datasets and the focus on clone detection in Java applications are among the most critical challenges of the available studies. With the fast-growing of large codebases, source code similarity measurement and clone-detection algorithms' efficiency, accuracy, and scalability have become important factors.

The rest of the paper is organized as follows. First, in Section 2, we briefly explain key terms and concepts used in code similarity measurement and code clone literature and review different types of code clones. Afterward, Section 3 outlines the research methodology and the underlying protocol for the systematic literature review. Section 4 describes our high-level classification of the primary studies and reports the results of this systematic review. Section 5 discusses the challenge and opportunities in the field. The threats to the validity of our SLR are discussed in Section 6. Finally, Section 7 concludes the SLR and presents future work.

## 2   Background

This section discusses the problem of code similarity measurement, clone detection, the types of similarity and clones, the origins and primary reasons for code similarity, and the related surveys performed in the field of code similarity and code clones.

### 2.1   Code similarity and clone definitions

Source code similarity is a concept used to measure the similarity degree between a piece of code snippets with respect to the text, syntax, and semantics. Although there is no well-established definition for code similarity in general, typically, the similarity score for any two code snippets $c_1, c_2$ is measured with a real number $s$ where the higher the score is, the more similar $c_1$ and $c_2$ are. Yamamoto et al. [56] have defined software system similarity as the ratio of similar element pairs to the total number of element pairs in the two given systems considering a concrete similarity metric. Qinqin and Chunhai [57] have stated that the similarity of program source codes is the same as that of the software systems, and Yamamoto's definition can be used as a basis for the similarity measure of program source code.

It is important to differentiate between the task of measuring code similarity and the task of identifying similar or clone codes. Indeed, the former is a general concept, and the latter can be categorized as one of the applications of code similarity measurement. For instance, the most similar instances can be reported as cloned instances by using a *threshold* value to filter out the results of similarity measurement [1]. As will discuss in this survey, code similarity measurement provides a basis to address different problems, denoting a wide range of applications in software engineering. Nevertheless, code clone detection is the most prominent application of code similarity, which is the subject of many articles and tools in this field. The advancements in clone detection methods can be promoted other code similarity measurement applications. Hence, we emphasize definitions and types of code clones. Researchers have proposed many definitions for code clones [1], [2], [48], [58]. As one of the widely accepted definitions, Baxter et al. [1] define the clone codes as follows:

**Definition 1 (Clone codes):** *Two code snippets $c_1$, $c_2$ are clones if they are similar according to some definition of similarity.*

Code snippet pairs in Definition 1 are in the specific granularity level in terms of different programming abstractions, including the statement, block, method, class, package, or component in a program. Moreover, code similarity measurement can be performed in the scope of one or more projects depending on the intended application. When entities of multiple projects are investigated to measure similarity and find similar instances, so-called *cross-project* similarity measurement techniques are defined and used [23], [26], [59]. In practice, the result of code similarity measurement can be reported with a vector, including the ranked similarity of the given code snippet and all code snippets with the same entity level.

Definition 1 leads to the formation of different clone types according to the definition of the similarity measure. Indeed, two additional aspects, including a similarity type and a threshold, must be considered for the clone detection task compared to the general definition of code similarity measurement. The similarity threshold is used to indicate the nearness of clone instances together. The maximum similarity value for this threshold indicates the exact matching of the instances. Consequently, not all similarity measurement approaches and similarity measures are suitable for code clone detection tasks. In this paper, we target the general problem of code similarity measurement. However, in most cases, the proposed techniques are only comparable based on the clone detection concepts, such as clone types.

Code clones can be detected for one programming language or between several languages. The latter type of clone detection is also called *cross-language* clone detection (CLCD) [60], [61]. Cross-language clones typically occur when a codebase is prepared to use in different environments for portability. For instance, the main codebase of the game is actually the same while building a game for different operating systems, and there are slight differences for each operating system [48].

## 2.2 Code similarity and clone types

Automatic code clone detection, in its general theme, is a complex and multi-purpose problem. Previous studies have recognized four types of clones that aim to simplify the problem [3], [22], [48]. The similarity between two pieces of code can be viewed and measured from two dimensions, including the syntax of the source code (program text) and the semantics or behavior of the source code (program functionality) [48]. Three types of clones, namely type I (exact clones), II, and III (near-missed clones), are recognized regarding the former dimension [8], [48], [62]. The latter dimension is led to define the type IV of clones [48], [63]–[65] which is also called semantics clones. It should be noted that there is no single or generic clone classification, and the proposed clone detection methods may use their definitions of code clones and clone types [66]. The widely-accepted definitions of clone types according to the existing research [3], [8], [22], [48], [62]–[66] are summarized as follows:

- **Type I**: The code snippets are entirely the same except for changes that may exist in the white spaces and comments.
- **Type II**: The structure of the code snippets is the same. However, the identifiers' names, types, white spaces, and comments differ.
- **Type III**: In this type, in addition to changes in identifiers, variable names, data types, and comments, some parts of the code can be deleted or updated, or some new parts can be added.
- **Type IV**: Two pieces of code have different texts but the same functionality.

Figure 1 shows an example of code snippets for each of the four clone types. The initial code fragment, $CF_0$, shows a `for` loop with a simple `if-else` block. In $CF_1$, only the comments have been changed, which led to the type I clone. In $CF_2$, the comments and identifiers names have been changed, resulting in a code clone with type II. In $CF_3$, the statement `a=10*b` is added as a new code fragment, which leads to a type III clone. Finally, in $CF_4$ the same program has been rewritten with a completely new structure. It is evident that finding the clones with type II is more challenging than finding type I. Similarly, detecting clones with type III is more complicated than type II clones due to the existence of new code fragments. Clone type IV is the most difficult to identify with the static analysis of the programs. Investigating the origins of code clone creation helps develop automatic clone detection and code similarity approaches.

```
CF0: Initial code fragment (CF)
1  for (int i = 0; i<10; i++)
2  {
3      // foo 2
4      if (i%2 == 0)
5          a = b + i;
6      else
7          // foo 1
8          a = b - i;
9  }
```

```
CF1: Type I clone
1  for (int i = 0; i<10; i++)
2  {
3      if (i%2 == 0)
4          a = b + i; // cmt 1
5      else
6          a = b - i; // cmt 2
7  }
```

```
CF2: Type II clone
1  for (int j = 0; j<10; j++)
2  {
3      if (j%2 == 0)
4          a = b + j; // cmt 1
5      else
6          a = b - j; // cmt 2
7  }
```

```
CF3: Type III clone
1  for (int j = 0; j<10; j++)
2  {
3      // new statement
4      a = 10 * b;
5      if (j%2 == 0)
6          a = b + j; // cmt 1
7      else
8          a = b - j; // cmt 2
9  }
```

```
CF4: Type IV clone
1  int i = 0
2  while (i < 10) {
3      // a comment
4      a = (i%2 == 0) ?
5  b+i : b-i;
6      i++
7  }
```

Figure 1. Examples of various types of code clones proposed in the literature [22].

## 2.3    Origins and root causes of code similarity and clones

The clone type categorization proposed in Section 2.1 helps design the different approaches suitable to detect specific clone types. However, it cannot reveal the origin and reasons for code clones. Various reasons lead to cloned codes in software systems during the life-cycle [42], [48]:

1. **Development strategy**: Specific development approaches such as copy-paste programming to reuse different software components create clones. For instance, generating code with a tool or merging two similar systems' source code often leads to clones.
2. **Using maintainability advantages**: Developers often make changes that increase code similarity during the maintenance phase.
3. **Programming language limitation and vendor lock-in**: The lack of built-in capability and abstraction mechanism for some widely used programming structures pushes developers to repeat specific pieces of code. For instance, the lack of inheritance and templates in the C programming language may lead to repeated same blocks of codes with minor changes. Such repeating structures may create potentially frequent clones.
4. **Accidental similarity**: Code similarities may be created inadvertently. For example, protocols for interacting with APIs and libraries typically require a series of function calls or sequences of commands that lead to code similarity. Moreover, two or more independent developers may use the same logic to solve a particular problem.

## 2.4    Related research

To the best of our knowledge, no systematic literature review has been performed on the general and particular aspects of the source code similarity measurement techniques and their applications, tools, and datasets. However, several studies have reviewed clone detection approaches [48], [67]–[71] or their specific applications [72]. For example, Novak et al. [72] have reviewed plagiarism detection in source code. Our survey complements the existing surveys with the recently proposed approach and emerging application of code similarity measurement without emphasizing a specific application. The goal of our SLR is to demonstrate code similarity measurement as a fundamental task underlying the automation of different software engineering activities.

Roy and Cordy [48] have published the first comprehensive technical report on software clone detection research. The authors have described the terminology frequently used in clone literature and mapped it to the commonly used clone types. They have also proposed a list of 57 open questions and issues on clone taxonomies, detection methods, and experimental evaluations. Some basic comparisons are made between clone detection methods and tools that provide valuable information for software developers. However, they did not use a systematic search to find and cover all related works, and their report was published nearly one decade ago.

Rattan et al. [67] have performed a systematic review of software clone detection and management. They have reported the result of reviewing 213 articles from 11 leading journals and 37 conferences. Their survey covers almost all software clone topics, such as clone definitions, clone types, clone creation reasons, clone detection methods, and clone detection tools. They have also tried to answer the fundamental question of whether the clone is "useful" or "harmful" with their reviews. Their review ended up with different and contradictory answers and concluded that simply deleting the clone instance is not a decent approach. Instead, developers require tools to identify and control clones to deal with the appropriate strategies in different situations. The authors have not reviewed the code similarity measurement in general.

Similar surveys on software clone detection have been recently performed by Min and Ping [68] and Ain et al. [69]. These studies cover approaches, tools, and open-source subject systems used for code clone detection, which help the researchers choose appropriate approaches or tools for detecting code clones according to their needs. However, the general problem of code similarity measurement and the wide range of its applications have not been covered systematically.

Concerning other applications of code similarity, the plagiarism detection techniques and tools have been reviewed by Novak et al. [72]. They have conducted a systematic review of the field to assist university professors in detecting plagiarism in source code delivered by students. Their study discusses definitions of plagiarism, plagiarism detection tools, comparison metrics, obfuscation methods, datasets used in comparisons, and algorithm types. The authors have focused on source-code plagiarism detection tools in academia and plagiarism obfuscation methods typically used by students who committed plagiarism. They have concluded that there are many definitions of "source-code plagiarism" both in academia and outside it, but no agreed clear definition exists in the literature. In addition, few datasets and metrics are available for quantitative analysis. We observed a similar status for other code similarity measurement relevant applications.

Some studies have empirically evaluated and compared a subset of source code similarity measurement and clone detection techniques [43], [62], [73], [74]. However, the results of such compassion are highly biased by the dataset. Burd and Bailey [73] have evaluated the performance of five code clone detection tools over a medium size Java program. They have concluded that there is no single and outright winner technique for clone detection. However, the reported results are only based on clones in a single program.

Bellon et al. [62] have evaluated the precision, recall, and execution time of six clone detectors using a benchmark containing eight large C and Java programs. The selected tools were based on different code similarity measurement approaches, including techniques that work on textual information, lexical and syntactic information, software metrics, and program dependency graphs. The authors have made references to code clones in eight software systems and compared the tools with their proposed dataset. They have concluded that the text-based and token-based tools demonstrate a higher recall but lower precision compared to tree-based and graph-based tools. In addition, the execution time of tree-based techniques is higher than other text-based and token-based. The paper has only used one human expert to judge the results of clone detection with threats to the reliability of the reported results.

Biegel et al. [74] have empirically compared the recall and the computation time of text-based, token-based, and tree-based source code similarity measurement approaches on four Java projects. They have used the tools to detect refactoring fragments in the code. The authors have found that the results of different tools have a large overlap. However, CCFinder [2], a token-based clone detection tool, is much slower than the other two similarity measurement approaches. The dataset and replication package of their empirical study is not publicly available.

In a recent empirical study, Ragkhitwetsagul et al. [43] have evaluated 30 code similarity detection techniques and tools. The authors have found that in the case of pervasive source code modifications, general textual similarity measurement tools offer similar performance to specialized code similarity measurement tools. However, in total specific source code similarity measurement techniques and tools can perform better than the general textual similarity measures. JPlag plagiarism detection tool [10] reveals the best performance on detecting local code changes, and CCFX achieves the highest performance on pervasively modified code. It should be noted that optimal parameters of the tools, such as the threshold used to detect clone codes naturally biased towards the particular dataset.

Chen et al. [70] have performed a critical review of different code duplication approaches between 2006 to 2020. The authors have concluded that there are no valid test datasets with a large number of samples to verify the effectiveness of available code clone detection. Moreover, the authors have reported that most techniques are highly complex, support a single programming language, and cannot detect semantic clones (type IV clones). However, they do not have investigated the wide range of code similarity measurement applications discussed in our paper. In this SLR, we review all the applications related to source code similarity measurement and discuss each of them thoroughly. We also expand existing categorizations of code similarity measurement algorithms and applications to encompass the most recent approaches taken by researchers and practitioners. Specifically, we added and discussed learning-based, test-based, image-based, and hybrid algorithms in addition to previously identified source code similarity measurement techniques. We also identified some emerging applications of code similarity measurement, including code recommendation and fault prediction.

# 3    Research Methodology

Our literature review follows the guidelines established by Kitchenham and Charters [75], which decomposes a systematic literature review in software engineering into three stages: planning, conducting, and reporting. Furthermore, we have taken inspiration from the most recent high-quality systematic literature reviews related to empirical software engineering, code smells, and software refactoring, software plagiarism [14], [76]–[78]. Figure 2 shows the overall process of research methodology, including the definition of the topic and research questions, search string creation, and article selection. The numbers inside the parenthesis are the number of articles retrieved at each step. We commence our research methodology by describing the research question that used to be answered during the SLR.

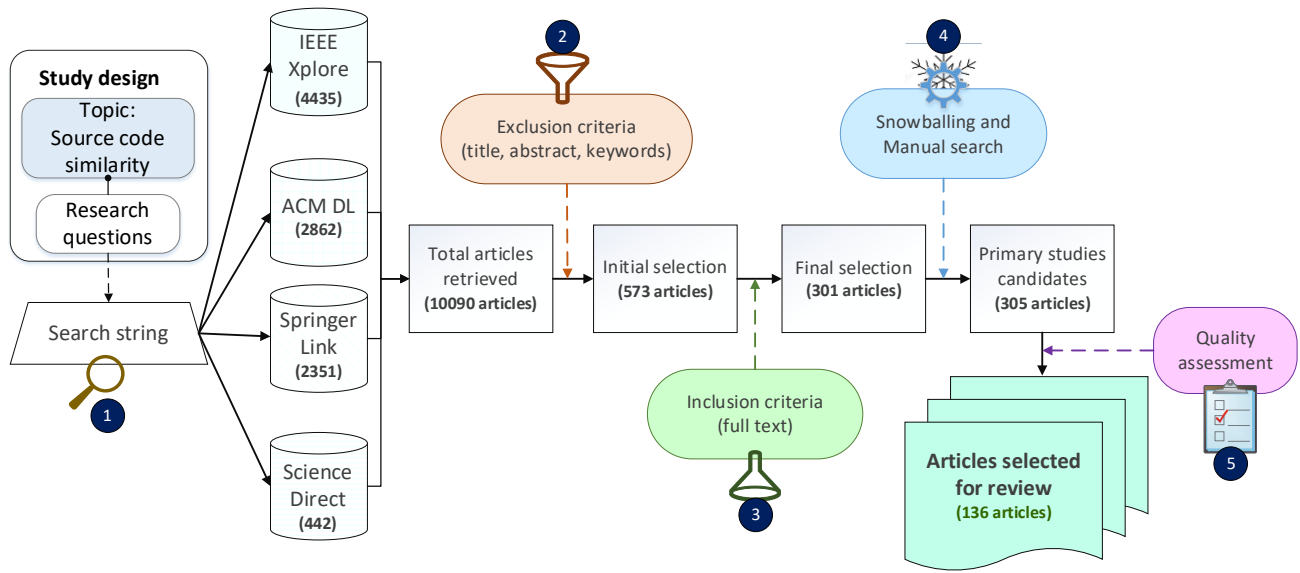

Figure 2. The research methodology used in our SLR.

## 3.1    Research questions

Source code similarity measurement and evaluation is a multidimensional problem with a broad range of applications. The following research questions were designed to conduct our review to find the objectives discussed in the introduction:

- **RQ1**: *What are the existing code similarity measurement approaches and techniques, and what are their advantages and disadvantages?*
- **RQ2**: *What are the applications of code similarity measurement in software engineering activities and tasks?*
- **RQ3**: *What are the existing code similarity measurement tools, and which programming languages are supported by each tool?*
- **RQ4**: *Which software projects and datasets have been used to build and evaluate code similarity measurement tools?*
- **RQ5**: *How is the performance of the existing code similarity measurement methods?*
- **RQ6**: *What are the existing challenges and opportunities in source code similarity measurement?*

## 3.2    Search string and digital libraries

After a deep investigation of related research in Section 2.4, we listed the most relevant keywords and terms considering the survey topic, *i.e.*, code similarity and research questions. The synonyms of collected words were also extracted and added to this list. Afterward, the logical operations 'AND's and 'OR's were used to correctly combine the terms and form a comprehensive search string, shown in Figure 3. Before performing the final search, we ensured the correctness of our search string by examining it on different search engines.

We searched the prepared string on the four well-known digital libraries, shown in Figure 2, indexing the majority of computer science and engineering publications. The libraries include IEEE Xplore, ACM Digital Library (ACM DL), SpringerLink, and ScienceDirect. These digital libraries index the most important journals and conferences in the field. Moreover, the Google Scholar library was used during the snowballing process. The initial number of articles obtained from each digital library has been shown inside the parenthesis in Figure 2. The search process was performed between April and May 2023, and studies published until that date were retrieved.

---

("code similarity" **OR** "similar code" **OR** "code clone" **OR** "clone code" **OR** "cloned code" **OR** "duplicate code" **OR** "duplicated code" **OR** "code duplicate" **OR** "replicate code" **OR** "code replicated" **OR** "clone detection")

**AND**

("measure" **OR** "vector" **OR** "metric" **OR** "detect" **OR** "detection" **OR** "identify" **OR** "compare" **OR** "comparison" **OR** "tool" **OR** "data" **OR** "dataset" **OR** "evaluate" **OR** "evaluation")

**AND**

("application" **OR** "technique" **OR** "approach" **OR** "method" **OR** "methods" **OR** "applications" **OR** "approaches" **OR** "techniques" **OR** "algorithm" **OR** "algorithms")

**AND**

("source code" **OR** "program" **OR** "code fragment" **OR** "code fragments" **OR** "software")

---

Figure 3. Search string used in code similarity measurement SLR.

It should be noted that when searching different digital libraries, we did not use any additional search filter for restricting the search to the title, keywords, or abstract of the articles, and the full text was considered in all databases. However, the necessary changes in the setup of our search string were applied due to the specific format imposed by the digital libraries. More specifically, we could not use the above search string with the ScienceDirect search engine because the engine has a limitation of accepting search strings including up to eight logical operators. Therefore, we had to look for the most relevant keywords in our search string to trim the query. As a result, the following logical expression was exclusively selected to search with the ScienceDirect library:

("Code similarity" **OR** "Source code similarity" **OR** "Code comparing" **OR** "Source code comparing" **OR** "code clone") **AND** ("Code clone" **OR** "Duplicate code" **OR** "detecting similarity" **OR** "clone detection")

## 3.3 Article selection process

As shown in Figure 2, a total of 10090 articles were initially found, which indicates that the code similarity measurement has received substantial attention. A systematic literature review should be based on influential and original articles in the field, called primary studies. We manually investigated the title, abstract, keywords, language, number of pages, and publication type to filter irrelevant articles and determine the primary studies. The subsequent sections explain the exclusion and inclusion criteria used in the article selection process.

### 3.3.1 Exclusion criteria

Some general exclusion criteria were automatically applied to filter irrelevant results using digital libraries' search engine features. On the IEEE Xplore library, we selected "Conferences" and "Journals" as the publication types to filter out irrelevant publications such as standards and books. On the ACM Digital Library, we selected "Research Article" for "Content Type" to exclude irrelevant publications such as posters, abstracts, keynotes, and works in progress. On the SpringerLink search engine, the "Computer Science" and "Engineering "disciplines were selected, and "Language" was set to "English". In addition, the "Content Type" filter was set to "Conference Paper" or "Article" to avoid retrieving non-relevant types of publications, mainly thesis and books. Finally, on the ScienceDirect library, we executed the trimmed version of our search string to retrieve relevant articles.

In addition to the general exclusion criteria, six specific exclusion criteria (ECs) were applied to filter out the redundant, irrelevant, and inappropriate articles from our SLR. The exclusion criteria used in our SLR are as follows:

- **EC1:** The articles repeated in citation databases were removed. We kept only one version of each duplicated article.
- **EC2:** Articles whose main text was written in any language other than English or only their abstract and keywords were in English were eliminated.

- **EC3:** Articles whose total text was less than three pages were removed. After reviewing them, we removed these articles and ensured they did not contain significant contributions. If there are good clues in these articles to find other suitable topics and articles, we consider these clues while applying the inclusion criteria and snowballing phases.
- **EC4:** Articles that did not directly discuss a source code similarity measurement approach in their abstract were removed. For example, some papers have discussed binary code similarity.
- **EC5:** The papers that had not proposed an *automated* approach for source code similarity measurement were removed. We excluded these articles since the similarity measurement technique was necessary when classifying methods.
- **EC6:** Theses, books, journal covers and metadata, secondary, tertiary, empirical, and case studies were removed.

Applying the above six specific exclusion criteria resulted in the deletion of 9517 publications, which were mostly irrelevant to the topic under review.

### 3.3.2    Inclusion criteria

It is challenging to eliminate an article from the remaining set without a complete study of the article. Therefore, we carefully applied a list of four inclusion criteria (ICs) to the candidate papers that survived in the previous steps. If all the following criteria were satisfied, we marked the article as a primary study:

- **IC1:** At least one keyword of the article existed in our search string.
- **IC2:** The title or abstract of the article is about the measurement or application of the source code similarity.
- **IC3:** The article has been cited at least once. This criterion was considered only for articles that were published before 2021. Using this condition, we could ensure other researchers in the community have considered the paper.
- **IC4:** The conference paper is considered if it does not have any journal extension. Otherwise, only its journal extension was included since articles published in a journal often offer more complete descriptions than their conferences counterpart.

Applying the above inclusion criteria resulted in 301 articles that constitute the candidate set of primary studies.

### 3.3.3    Snowballing and manual search

To ensure the finding and include all relevant articles in our primary studies, we performed both forward and backward snowballing [79] on the set of 301 studies. In backward snowballing, the references of each paper are searched in the articles initially found by searching digital libraries. Any reference which has not existed in that list is added to a new list. In forward snowballing, citations of each paper are investigated to find that the relevant paper has not been found already. As shown in Figure 2, the snowballing process only found four new articles, resulting in 305 candidate articles. It demonstrates that our search string is well-designed such that it can find almost all relevant articles in the field. Some articles published at conferences have not been indexed in the four well-known digital libraries. These articles were found during the snowballing process in Google Scholar.

## 3.4    Quality assessment

We assessed the quality of the remaining articles using a standard quality assessment procedure by Kitchenham and Charters [75]. The following checklist has been designed to evaluate the quality of the papers that passed our inclusion and exclusion criteria:

- **Q1:** *Does the paper propose a new approach to source code similarity measurement, clone detection, or their applications?*
- **Q2:** *Does the paper specify the types of code similarity or clones it detects?*
- **Q3:** *Does the paper offer any software tool for source code similarity measurement or clone detection?*
- **Q4:** *Does the paper use any dataset for evaluating its proposed method?*
- **Q5:** *Are the paper's findings clearly stated and compared to the related works?*

Each of the above checklist questions was marked with "Yes," "No," or "Partially" for each paper in the candidate list of primary studies. The answers were scored based on the following three rules: "Yes" = 1, "No" = 0, and "Partially" = 0.5. The quality score of each candidate's primary study was calculated as the sum of the scores obtained by all the questions.

The scoring process was performed by the first, third, and fifth authors, who jointly evaluated each candidate's article and used consensus to determine the final score. The second and fourth authors were asked to resolve conflict and made the final decision when consensus was difficult. We categorized the article's quality level into High ($4 \leq score \leq 5$), Medium ($2.5 \leq score < 4$), and Low ($score < 2.5$). The articles whose scores belonged to the High and Medium levels were selected for in-depth analysis as our final primary studies.

During the quality assessment process, 169 articles were removed, and 136 articles were selected as the primary studies about code similarity measurement and evaluation. We observed that articles without in-detail and helpful information about their methods and evaluations received a relatively low score and were removed from the primary studies. The complete list of primary studies has been mentioned in Appendix A. We used Microsoft Excel to organize the results of the article selection process and handle the information extracted for each primary study. The Excel file containing the full list of retrieved papers and the filter results is publicly available on Zenodo [80].

## 4    Results and Findings

This section discusses our findings and the result of reviewing the primary studies obtained in Section 3. We commence the section with a brief demographic of statistical data obtained from our systematic search.

### 4.1    Demographics

Figure 4 shows the number of studies published on source code similarity measurement and clone detection each year. For better visualization, all publications before 2005 are shown in the same bar as the year 2005. Initial studies include all remaining studies after applying exclusion criteria, and primary studies include 136 articles in the final selection. Overall, we observe a growth rate in the number of publications in this field in recent years, indicating an increase in the need for code similarity measurement approaches and tools. Two spikes are observed in the number of primary studies in 2013 and 2020, respectively. However, the number of studies dropped in 2021, possibly due to the long-term impacts of the COVID-19 pandemic. Some papers written in these years possibly have not been published at the time we searched for the literature. Table 1 shows the publication type of primary studies which have been indexed in each digital library. One paper had not been indexed by any of the four digital libraries we searched and it was found in Google Scholar. About 27.94% of the primary studies have been published as journal articles, and the remaining 72.06% are conference papers.

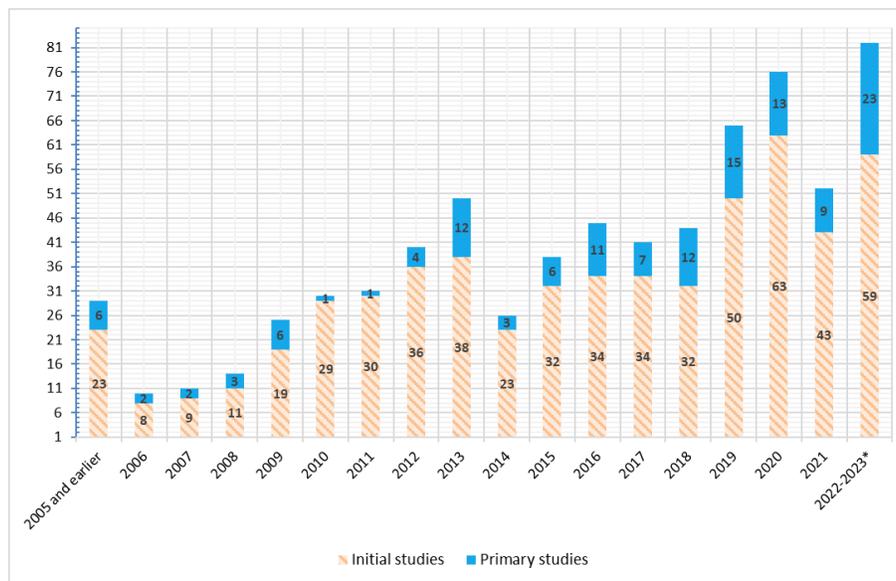

Figure 4. Number of initial and primary studies per year.

Table 1. Type of publications in code similarity measurement.

| Database | Journal | Conference | Total |
|---|---|---|---|
| IEEE Xplore | 14 | 59 | 73 |
| ACM Digital Library | 1 | 26 | 27 |
| Springer Nature | 11 | 13 | 24 |
| ScienceDirect | 11 | 0 | 11 |
| Google Scholar | 1 | 0 | 1 |
| Total | 38 (27.94%) | 98 (72.06%) | 136 |

## 4.2 Classifying and comparing code similarity studies

The analysis of the primary studies indicates that code similarity measurement studies expose some common aspects. Appropriate classification of these aspects provides many benefits for comparing different code similarly measurement approaches in a standard scheme and disclosing the research gaps in the field. We designated five orthogonal dimensions for each study about the code similarity measurement, including technique, tool, supported programming language, dataset, and application domain. Figure 5 demonstrates our proposed classification and the existing subcategories in each category. We organized our survey based on this classification. The proposed classification can be considered a standard way to systematically compare the current and future research on code similarity. Unfortunately, not all studies provide information and discussion in all five proposed categories, shown in Figure 5. In the rest of Section 4, we offer an in-depth review of each of these five dimensions to discover the current state of research in code similarity measurement.

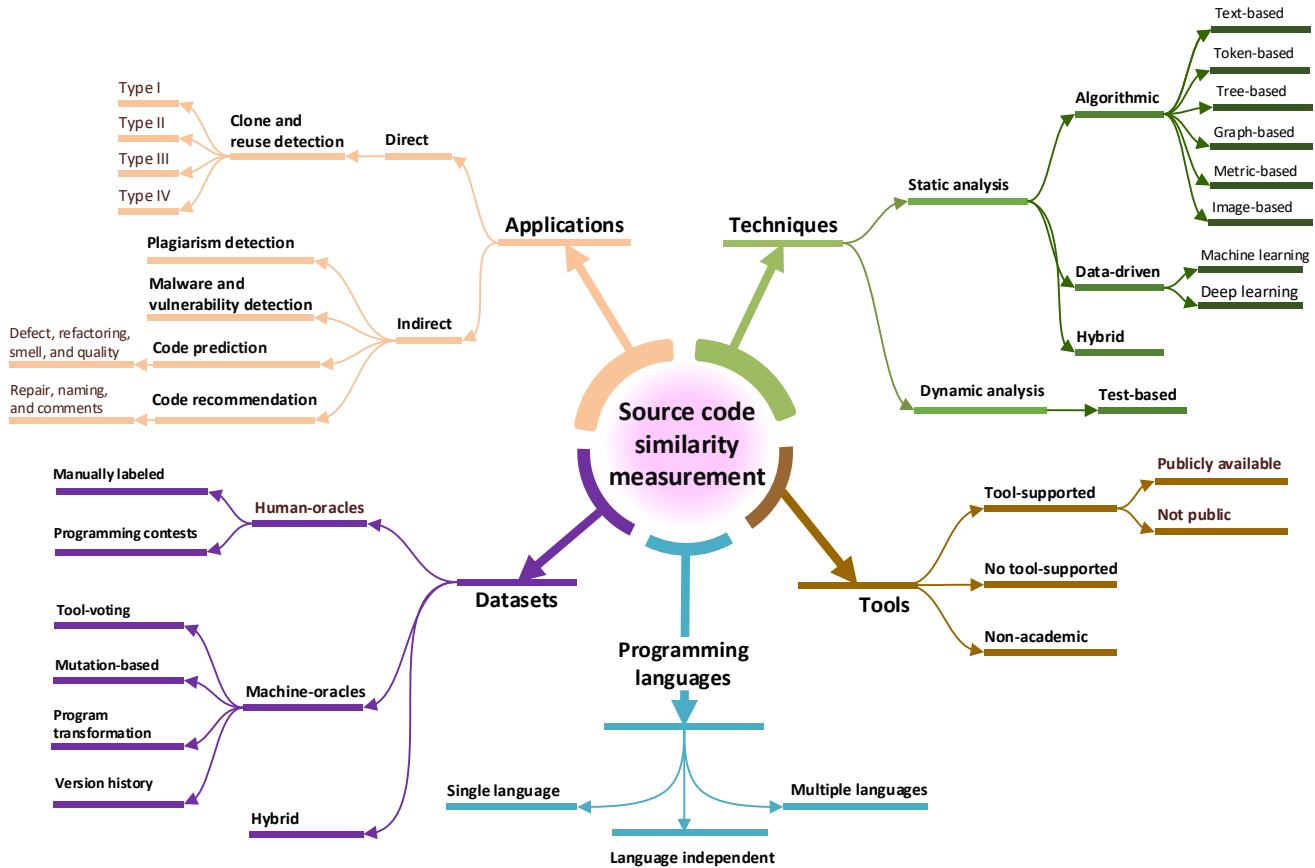

Figure 5. Classification of different aspects in the code similarity studies.

Regarding technique, the existing studies are either based on static or dynamic program analysis. Methods based on static analysis such as [10], [40], [61], [81]–[83] do not use any execution of the programs to measure similarity. They are using algorithmic, data-driven, or hybrid approaches to find similar source codes. Few studies [84], [85] use dynamic analysis in which programs are executed with a test suite or working load to obtain runtime information such as program function outputs and then find similar code fragments by comparing the obtained information. Section 4.4 discusses existing techniques in detail.

Regarding the tools, proposed studies may be supported by a tool or not. Some studies have publicized their tool [10], [81] while others have not. There are also non-academic code similarity measurement tools that have not been discussed by primary studies [86]–[89]. The existing tools and approaches may detect source code similarity in one or multiple programming languages. There are also language-independent tools [90]. Section 4.5 describe existing code similarity measurement tools and supporting languages.

Our review of primary studies shows three kinds of datasets used to create and validate code similarity measurement methods: Datasets with human oracles [62], [91], datasets with machine oracles [92], [93], and hybrid datasets [43]. The first category of datasets is either manually labeled by experts in the field or is based on the solutions submitted to specific problems in programming contest platforms such as Google CodeJam [94] and Codeforces [95]. Automatically generated dataset uses majority voting between different existing tools [62], mutation operators [93], program transformation rules [96], or version history of a codebase to make labels for similar and non-similar code fragments without human intervention. Finally, a combination of these approaches can be used to create large and quality datasets. The details of existing code similarity benchmark and datasets are discussed in Section 4.6.

Regarding the applications of code similarity measurement, we observed a direct application of code clone and reuse detection and several indirect applications, which were further classified into plagiarism detection, malware, and vulnerability detection, code prediction, and code recommendation. While the majority of primary studies have focused on finding clone instances, our survey indicates that source code similarity measurement can address many code-related tasks in software engineering activities. Section 4.3 describe the different applications of code similarity measurement covered by the studies we reviewed.

## 4.3    Code similarity measurement applications

Our SLR indicates that in addition to code clone and plagiarism detection, there are three other code similarity measurement applications, including malware detection, defect prediction, and code recommendation. Indeed, the three later applications have rarely been discussed from the code similarity measurement viewpoint. Figure 6 illustrates the frequency of the top five code similarity measurement applications in primary studies. According to the pie chart in Figure 6, code clone detection is the most frequent application of code similarity measurement research. It has been discussed in over 84% of the primary studies. Indeed, most studies have investigated code similarity in the general domain of code clones instead of directly talking about a specific application. Code clones are detected in both production and test codes [97].

Code plagiarism detection is one of the trivial code similarity measurement applications with the second frequency rank in which the goal is identifying unauthorized reuse of the program source code [72], [98], [99]. In malware detection, the third frequent application of code similarity measurement according to our SLR, the program source codes are searched to find clone instances of malicious source codes suite [17]–[19], [100]. These instances most probably contain malicious behavior. Similarly, clone instances of a source code suite with known defects are identified as programs that possibly have flaws [101]–[103]. Finally, code recommendation [26], [104] is one of the most recent applications of code similarity. It recommends code modification and improvements such as bug-fixing patches, entity names, refactorings, and code generation based on the similarity of the developing code to the existing code in a given corpus. For example, several approaches have been proposed to suggest methods and class names to the programmers to improve the source code quality factors [23], [24], [26], [105], [106].

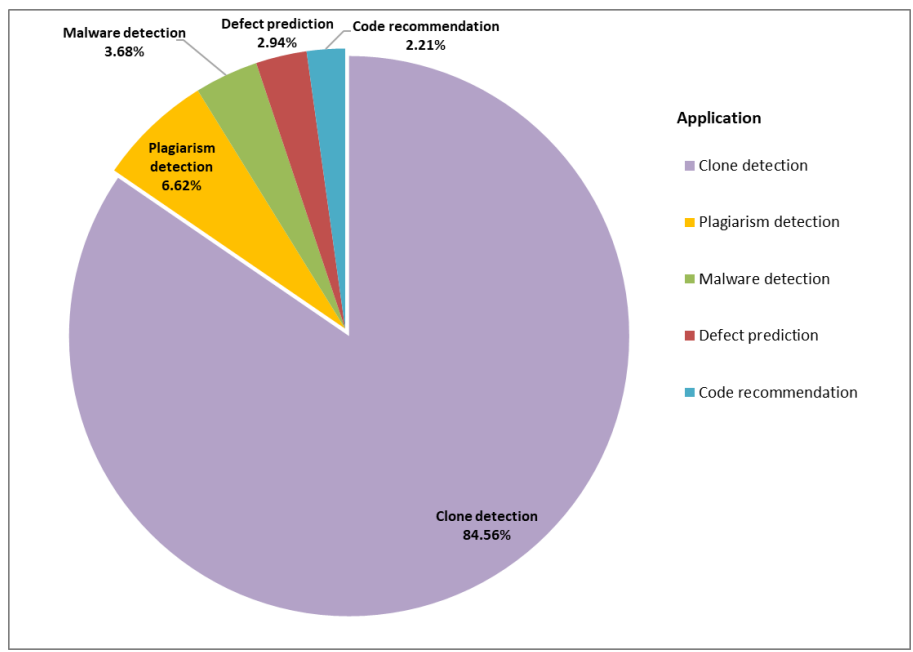

Figure 6. Source code similarity measurement applications.

## 4.4    Code similarity measurement techniques

Many methods and techniques have been proposed in the literature to identify similarities, while few have been commercialized. Indeed, most approaches have been developed in academic research with the purpose of software maintenance. We identified nine techniques to measure code similarity used by the primary studies. Figure 5 illustrates the structured levels of source code similarity measurement techniques. Previous surveys have considered at most six categories in the leaves of our structured tree, but our SLR could identify three new categories of techniques for source code similarity measurement, including learning-based, image-based, and test-based methods.

Figure 7 illustrates the frequency of different code similarity measurement techniques in primary studies. It is observed that hybrid methods, which combine two or more base methods, are the most used techniques for code similarity measurement. According to the figure, nearly 28% of studies have proposed a hybrid method to measure source code similarity. Learning-based and token-based methods are in the second and third ranks, respectively. The graph-based, tree-based, and text-based techniques are respectively in the fourth, fifth, and sixth ranks. Finally, the metric-based, test-based, and image-based methods are the least examined methods. The most important primary studies proposed based on each technique are discussed in the rest of this section to provide the mechanisms and details of each method.

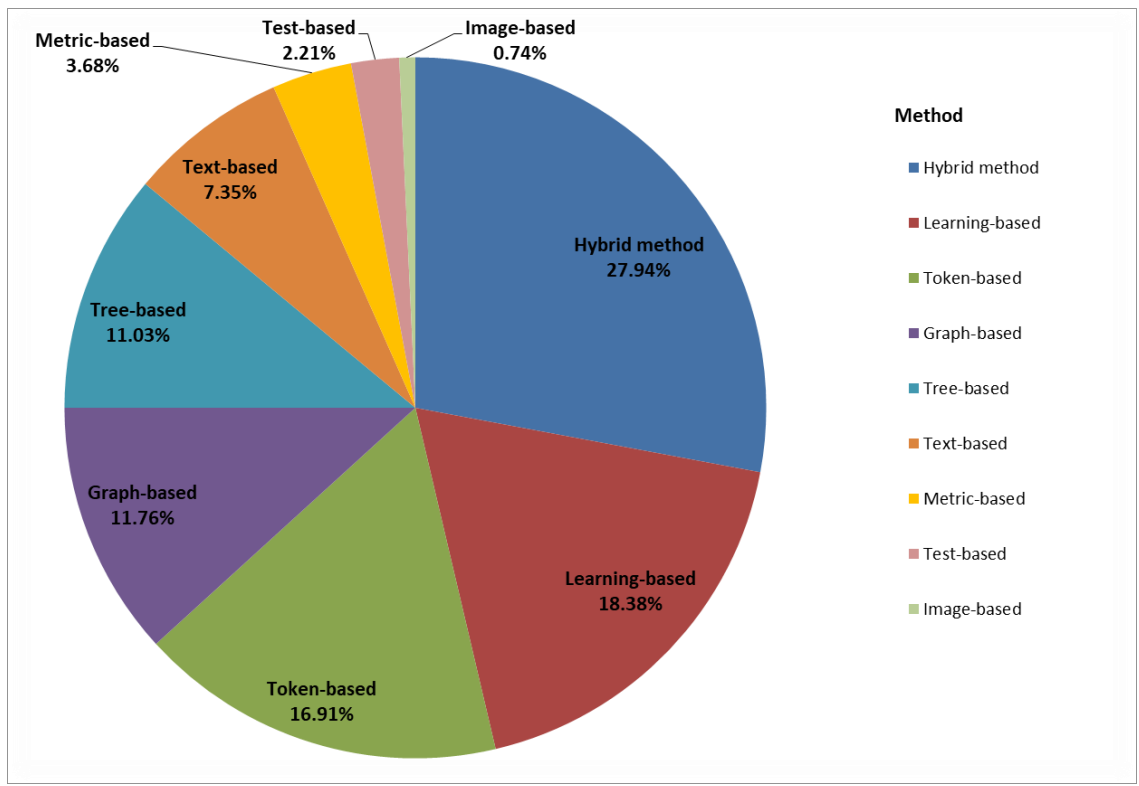

Figure 7. Frequency of different code similarity measurement methods.

### 4.4.1    Text-based techniques

The simplest and early-devised technique for code similarity measurement is to treat the source code as a text document [P2]. Text-based methods consider the source code as a sequence of strings [P30]. Two code snippets are compared together to find the longest common sequence of strings then matching parts is reported as clone instances. The comparison is typically performed on raw codes without any preprocessing. In some applications, the white spaces and comments are removed before sequence matching. The main advantage of text-based methods is that they are efficient and can be applied to any programming language without any modification.

Burrows et al. [P10] have proposed a text-based approach to finding plagiarism in source code. The similarity between sequences of string obtained from the source code is computed using the local alignment procedure [107], an approximate string-matching algorithm. To achieve better efficiency each token in the evaluated corpus is stored and indexed only once. The authors have shown that the effectiveness of their approach is comparable to the JPlag [P6] tool while it is highly scalable and efficient due to filtering and indexing the source code fragments. However, determining an appropriate threshold for the local alignment algorithm to select similar instances is challenging.

NICAD is a well-known and most-used text-based clone detection tool that uses text normalization to remove noise, standardize formats, and break program statements into parts such that potential changes are detected as simple line-wise text differences [P11], [P14]. Ragkhitwetsagul and Krinke [P63] have used compilation and decompilation as pre-processing steps to normalize syntactic changes in the code fragments and improve the accuracy of NICAD. gCad [P35] is a near-miss clone genealogy extractor based on NICAD detecting clones in multi-versions of a program. It accepts $n$ versions of a program, maps clone classes between the consecutive versions, and extracts how each clone class changes throughout an observation period.

Cosma and Joy have [P22] proposed a text-based approach to detecting plagiarism supported by a tool, PlaGate. They have used the latent semantic analysis (LAS) technique to convert source code files to a numerical matrix and then compute the cosine similarity between each pair of files in a corpus. PlaGate is language independent source code plagiarism detection tool. However, choosing the right dimension for the LSA matrix on each corpus is challenging.

Kamiya et al. [P45] have proposed a more advanced method in which a lexical analyzer creates a sequence of text. Then the AST is constructed for each sequence, and common subtrees are found using the Apriori algorithm [108]. Their method finds the type I and II clones with sequence matching and type III clones with AST matching. The advantage of this method is that it does not need to create the AST for the entire program. However, it does not support detecting clones of type IV.

Chen et al. [P43] have used a text-based method to identify malware applications for Android systems. The source code is broken into small code snippets, and the similarity between these snippets and existing malware is computed using the NiCad tool [109] to detect the malware. The score of this method is over 95% (with a recall of 91% and precision of 99%), which is relatively high compared to the other text-based methods.

Tukaram et al. [P88] have proposed another text-based method that uses a preprocessing module to remove extra information in the code, such as spaces. Other modules used in their approach include keywords, data types, variables, and functions detector, identifying a specific part of the code. Finally, it uses a Similarity Checker module responsible for identifying similarities. This article aims to provide general information on the similarity of applications' codes at different levels. The advantage is that it can be used to evaluate the software and give general-purpose details about the clone code.

### 4.4.2 Token-based techniques

In token-based methods, the source code of programs converts to sequences of tokens, and then these sequences are compared to find common subsequences [P7], [P98]. Token-based techniques are more robust to code changes, such as changes in identifier lexemes, than text-based methods [P15], [P18], [P26]. They are also efficient enough to be used frequently during the software development process [P53]. However, the processing time increased due to token recognition and replacement operations. Token-based methods have demonstrated a good performance in identifying code clones with type I and type II [P13], [P87], [P89], [P92], [P95], [69] as well as in detecting large-gap clones which are type III clones with many edits [P68], [P75], [P79].

Rehman et al. [P25] have developed a tool, LSC Miner, to find clone codes with a token-based method. LSC Miner supports clone detection for multiple languages, including C, C++, VB6, VB.NET, and Java. Misu et al. [6] have stated that two functions with similar interfaces are prone to clone. Their approach first creates the corresponding interface for each method in the program. Then, the Jaccard similarity [110] between the sequence of tokens for each pair of methods is computed. The sequence includes a token stream of function arguments and identifiers. Function with a similarity of more than 70% is grouped as clones' instances. This threshold has been determined experimentally. They have developed their method for Java programs. Ankali and Parthiban [P113] have used the Levenshtein distance to search cross-language code clones in Java, C, and C++. The classification of clone types is performed manually by the authors.

Lopes et al. [P61] have detected clone code in GitHub repositories with a token-based method on C++, Javascript, Java, and Python programs. To this aim, they have used the SourcererCC tool [P47] to check the source codes at the project and file levels. They concluded that about 14% of Java, 25% of C++, 18% of Python, and 48% of JavaScript codes have clone instances in GitHub repositories. CP-Miner [P7] finds copy-paste and their related bugs in large-scale software code.

JPlag [P6] tokenizes the program as a sequence of strings and uses the Greedy String Tiling [111] algorithm, which finds a maximal set of contiguous substrings. It provides a web service that finds pairs of similar programs among a given set of programs. JPlag [P10] is mainly used for detecting source code plagiarism. Liu et al. [P79] [112] have used the Smith-Waterman sequence alignment algorithm for token matching to increase the efficiency of token-based clone detection. CPDP is a similar token-based tool for plagiarism detection in source code [P32].

Source code plagiarism detection systems can be confused when structural modifications, such as changing control structures, are applied to the original source code [P32]. To mitigate such issues, Duric and Gaševic [P26] have proposed an enhanced token-based plagiarism detection tool, SCSDS, with specialized preprocessing that avoids the impacts of some structural modifications. For example, all identifiers of Java numeric types, *e.g.*, int, long, float, and double types, are substituted with the $< NUMERIC\_TYPE >$ token. In addition, all semicolon tokens are removed. SCSDS achieves a better F1 score compared to the well-known JPlag tool [10]. However, the comparison has been performed on a tiny dataset of students' programming assignments which threats the generalization of the approach.

Ullah et al. [P75] have used a token-based clone detection to detect plagiarism in students' programming assignments. Their method first creates a frequently repeated program tokens matrix using latent semantic analysis (LSA) [113]. This method helps convert the source code to Natural language. The similarity between programming assignments is computed by LSA [113]. The LSA algorithm has been applied to extract semantics from tokens to find plagiarism among the students' programming assignments in C++ and Java. The authors have reported a maximum recall of 80% for their proposed method.

CP-Miner [P7] finds both copy-pasted code and copy-pasted efficiently in large software suites. It first tokenizes the program and then applies a subsequence mining algorithm called CloSpan [114] to find the frequent subsequence between code fragments. The most frequent subsequences are exactly copy-pasted segments in the original program. Experiments with the CP-Miner tool show that about one-third of copy-pasted codes contain the modification of one to two statements. CP-Miner is a language-independent tool. However, its time complexity is $O(n^2)$ where $n$ is the maximum length of frequent sequences.

Detecting type IV code clones with text-based or token-based methods is challenging since the programs' text is often different. Rajakumari et al. [P89] have addressed the detection of type IV clones for Java programs in five steps: selecting input, separating modules, extracting clones, classifying clones, and reporting results. Ragkhitwetsagul et al. [P87] have used well-known TF-IDF techniques to convert the sequences of tokens to vectors. First, the indexing phase is executed to create the vector representation of the token and assign scores to each vector. Then, the detection phase selects similar instance and report them as clones. Their proposed tool, Siamese, can detect type I, II, and III clones for Java programs with a maximum recall of 99%. Another advantage of this method is its relatively high time performance in detecting clone instances. However, Siamese cannot detect type IV clones. FastDCF [P128] is another token-based clone detection tool that uses HDFS [115] and MapReduce [116] to scale up clone processing into big codebases at the granularity of the function and files. Similar to Siamese cannot detect type IV clones.

### 4.4.3 Tree-based techniques

The source code is converted to the parse tree or abstract syntax tree (AST) using a programming language parser in tree-based methods [P8], [P16], [P78], [P121]. The search for similar codes is performed in the parse tree or AST to find similar subtrees [P24]. This method is robust to code changes such as different blocking and paracentesis. However, creating the parse tree or AST for large codebases is time-consuming and requires a specific parser for each programming language [P93]. In addition, matching subtrees is computationally expensive [P93]. In the application of code clone detection, tree-based methods can recognize code clones of types I, II, and III accurately. They have been also applied to other application domains of code similarity measurement such as malware detection [P74] and defect prediction [P115].

DECKARD [P9] is a tree-based code clone detection that uses a tree edit distance metric to measure code similarity. The edit distance of two trees $T1$ and $T2$, denoted by $\delta(T1, T2)$, is the minimal sequence of the edit operations required to transform $T1$ to $T2$. The DECKARD [P9] performance highly depends on the threshold used to determine similarity. Gharehyazie et al. [P91] have developed the CLONE-HUNTRESS tool based on DECKARD to identify and track the clone instances across GitHub projects. However, their tool cannot capture type IV of clones. Chandran et al. [P57] augment the abstract syntax trees (ASTs) of functions considering security-relevant properties of the code to find vulnerable and buggy clone instances.

Tekchandani et al. [P27] have proposed a tree-based approach for detecting type IV clones. First, the source code AST is first created, and then the sequences of ASTs' leaves are compared to identify clone pairs. Their approach tries to extract grammar from the source code samples and build AST to support multiple programming languages. Therefore, this method works well in cases where a programming language's grammar is unavailable. However, the clone detection time increased due to extracting grammar from the source code samples, while its accuracy decreased due to errors in the grammar extraction process. Asta [P19] uses a similar approach to convert AST nodes to sequential patterns with a limited length and compares the patterns for detecting possible clones.

Gao et al. [P78] have proposed a tool, TECCD, which uses a word2vec algorithm [117] on the program AST to reduce the time of tree-based similarity measurement. The ANTLR parser generator [118] is used to create the source code AST, and a matrix containing AST nodes is fed to a word2vec model resulting in the vector representation of each AST. The Euclidian distance between pairs of corresponding AST vectors is computed as a measure of similarity. The learned model is used to convert new source code to vector, which improves the efficiency of the tree-based methods. It can detect code clones of type I, II, and III with a precision of 88% and recall of 87%. TECCD cannot detect type IV clones.

Sager et al. [P8] have tried to identify similar classes in Java code using the AST. Their method obtains the AST of each class using Eclipse's JDT API. Afterward, the AST is converted to an intermediate model called FAXIM (FAMOOS Information Exchange Model), an independent programming language model for displaying object-oriented source code [P8]. The similarities between the two classes are then compared using comparative algorithms, including bottom-up maximum common subtree isomorphism, top-down maximum common subtree isomorphism, and the tree edit distance. The similarity is measured between FAXIM trees. The discussed method identifies similar classes with an average recall of 74.5% on two software projects.

### 4.4.4    Graph-based techniques

Graph-based methods typically create the program dependence graph (PDG) [119] for code snippets and then compare these graphs to recognize the similarity [P21]. The PDG contains both the data and control dependencies between program statements and hence conveys the semantics of the program in addition to its syntax. The node of PDG are program statements, and the edges are data or control dependencies. The proper implementation of PDG-based methods can identify all types of clones [P34], [P119]. However, comparing PDGs is often made with graph isomorphism, which is an NP-complete problem [P23], [P31]. Therefore, some approaches convert PDGs into more simple tree representations for use in clone detection [P64].

Komondoor and Horwitz [P4] have proposed a graph-based clone detection approach to find clones whose components do not occur as contiguous text in the program. Their algorithm finds two isomorphic subgraphs for each pair of matching nodes in the PDG using a backward slicing technique. The matching nodes are nodes in the PDG whose statements are synthetically similar without considering the variable names and values. While the slice-based approach is more efficient than finding the isomorphic subgraphs in general, it is time-consuming for programs with large PDGs.

Krinke [P3] has proposed a graph-based approach to identify similar codes by calculating the maximal similar subgraphs. The approach can detect semantically similar codes. However, according to the authors, the time complexity of the proposed algorithm is $O(|V|^2)$ where $V$ is the number of PDG vertices. Therefore, for the code fragments with large PDG, the algorithm is not efficient. Gabel et al. [P12] have reduced the graph similarity measurement problem to a simpler tree similarity measurement problem by mapping the PDG subgraphs to their related structured syntax. Using tree similarity measurement make the process of comparing subgraph efficient.

Wang et al. [P58] have proposed a clone detection tool, CCSharp, based on PDG for C# programs. CCSharp breaks source code into small snippets and builds PDG for each code snippet to improve PDG creation time. The graphs are then compared together to detect and group clone instances. The advantage of CCSharp is detecting all clone types with a relatively high precision of 99.3% and recall of 89.8%. Tajima et al. [P71] have extracted interface information and PDG from the program's methods to use for similarity detection within the same project.

Constructing PDG for large programs is time-consuming and error-prone. Avetisyan et al. [P41] have proposed an LLVM-based code clone detection framework using program semantic analysis based on PDG. Their proposed clone detection approach is a multi-stage process. First, generated PDGs are loaded into the memory. Second, the PDGs are split into subgraphs with various techniques. The third step is the application of fast check algorithms which try to prove that the pair of PDGs does not have big enough isomorphic subgraphs to be detected with linear complexity. The fourth stage is the maximal isomorphic subgraphs detection based on program slicing. The fifth step is filtration which checks that the source code for the corresponding isomorphic subgraphs is not too scattered. Finally, the corresponding source code for isomorphic subgraphs is printed as detected clones. The authors have focused on the efficiency and scalability of the approach, and they have identified types I, II, and III of code clones. However, their approach cannot detect clone type IV and is restricted to the LLVM ecosystem [120]. Some approaches have used evolutionary algorithms such as the genetic algorithm to find similar subgraphs efficiently for detecting similar source code [P51]. However, these approaches find cloned code with approximate graph matching which may not be accurate [P110].

Another type of graph extracted from the programs to identify similarity at coarse grain is the function call graph. Gascon et al. [P33] have proposed a method for malware detection based on efficient embeddings of function call graphs and machine classification models. They have achieved an 89% of recall in detecting Android malware. However, the static analysis used for the call graph construction may not be accurate, specifically in the case of the polymorphic methods.

Marastoni et al. [P60] have proposed an approach based on static analysis, which works by computing the CFG of each method and encoding it in a feature vector used to measure similarities. Their tool, Groupdroid, aims to detect Android malware applications. The Groupdroid tool has achieved an F1 of 82%. However, the approach only works at the method level, while the malware interactions can spread among different methods. Kalysch et al. [P74] have used a similar approach by extending the extracted features from CFG to improve the effectiveness of Android malware detection. They have achieved a precision of 89%.

Kim et al. [P44] have proposed a graph-based method to identify malware applications for Android systems. The structural information of methods is extracted from methods in given applications and compared to match the similar methods of target applications, and the number of matched methods and the total number of methods are used for similarity calculation. The similarity calculation process consists of two steps. First, control flow graphs of each method in two Android applications are extracted. Secondly, structural information that is collected from the control flow graphs is compared to match the similar methods in two target applications. However, extracting and comparing CFGs for each pair of methods in two applications are time and resource-consuming.

### 4.4.5    Metric-based techniques

Metric-based methods first represent source code as a vector of different code metrics, using the vector space model theory [121]. Then, vector similarity is used to measure the similarity between each given source code. The vector similarity highly depends on the quantity and quality of defined metrics for the programming language. Computing source code metrics for large programs takes reasonable time resulting in the proper efficiency of this method [P1], [P114]. Table 2 shows the common source code metrics used for code similarity measurement and clone detection. It is observed that researchers use only a limited set of source code metrics.

Sudhamani et al. [P56] have proposed a metric-based clone detection approach for Java programs. Their method extracts a set of Control Feature Metric Tables (CFMT) for control statements in the program, including the number of arithmetic operations, logical operations, conditional operations, identifiers, and lines of control codes. A vector of extracted metrics represents the code snippets, and the Euclidian distance is used to measure the similarity between each pair of code snippets. The authors have claimed that their approach can detect all types of code clones. However, a precise evaluation of this method has not been reported.

Nafi et al. [P100] have used nine code metrics to detect cross-language clones (CLC). The ANTLR tool [118] has been used to extract source code metrics, and the cosine similarity distance has been computed as a similarity measure. A typical example of (CLC) is the development of mobile phone games, where developers try to make a single game available on different platforms. For instance, a popular mobile phone game, Temple Run, is available on three mobile phone platforms, Windows, iOS, and Android. The goal is to identify different clones between C#, Java, and Python programming languages. This method first obtains software metrics and then compares each piece of code with the metrics. The F1 score of this method in different languages has been reported to be between 65% and 71%.

Sudhamani et al. [P86] provide a metric-based method for identifying semantic clones. They used metrics such as the type and order of control statements, nested control statements, type and count of operators, and the number of variables. This method consists of five steps: 1. Pre-processing, 2. Computation of features, 3. Dissimilarity matrix computation, 4. Similarity value computation, 5. Similar program detection through clustering. The metrics are computed and placed in the Control Statement Feature Table (CSFT) matrix for comparison. The dissimilarity between different matrices is then calculated based on city block distance, and finally, the codes that are not similar are not deleted. The method's F1 varies according to the surveys performed on different Java projects, but on average, it is between 90% and 95%.

Parsa et al. [P133] have proposed a metric-based approach to find similar Java methods for the method name recommendation task. They have shown that a lightweight program representation based on source code metrics achieves better accuracy than complex deep learning-based models, such as Code2Vec [P85] and Code2Seq [24] in code-related tasks, specifically method name suggestion. However, their approach cannot detect semantically similar code fragments.

Table 2. Source code metrics used in code similarity measurement approaches.

| Metric name | Main references |
| --- | --- |
| Lines of code | [P133] |
| Number of variables declared | [P100], [P133] |
| Total number of operators | [P100], [P133] |
| Number of arguments | [P100], [P133] |
| Number of expressions | [P100] |
| Total number of operands method | [P100] |
| Number of loops (for, while, and do-while) | [P56], [P86], [P100], [P133] |
| Number of exceptions thrown | [P100] |
| Number of exceptions referenced | [P100] |
| Cyclomatic complexity | [P100], [P114] |
| Number of local variables | [P39], [P62], [P133] |
| Number of function calls | [P133] |
| Number of conditional statements | [P86], [P133] |
| Number of iteration statements | [P86], [P133] |
| Number of return statements | [P86], [P133] |
| Number of input statements | [P86], [P133] |
| Number of output statements | [P133] |
| Number of assignments through function calls | [P133] |
| Number of selection statements | [P133] |
| Number of assignment statements | [P133] |
| Number of passed parameters | [P133] |

### 4.4.6 Learning-based techniques

We preferred to discuss the learning-based method as a separate category, given the fact that machine learning has become widespread in recent years [122]. Machine learning methods, summarized in this section, have shown promising results in code similarity measurement and clone detection tasks [P17], [P42], [P49], [P59], [P84], [P101], [P102], [P103], [P108], [P109], [P118], [P120], [P132]. The ability to learn and recognize complex patterns made the learning algorithms suitable for these tasks. However, learning-based required large and clean code clone datasets to work correctly, which is not available for all languages. Most approaches use the existing code clone detection tool to prepare the required data for machine learning, which is unreliable since the tools always make errors.

Learning-based approaches aim to classify or cluster similar code fragments by learning over a given dataset of known similar and dissimilar codes [P42], [P59], [P84], [P102], [P105], [P107]. Cesare et al. [P37] have used conventional classification models to identify clones at the package level. The authors reported Random Forest as the best learning model. Joshi et al. [P46] have used the DBSCAN clustering algorithm [123] to identify types I and II of clones at the function level. Keivanloo et al. [P42] have used the $k$-means clustering algorithm to determine true and false clones of type III with any dissimilarity threshold, which is common in the non-learning methods. The authors have used the Friedman method to evaluate the clustering quality for different values of the $k$ parameter in the clustering algorithm. Their approach cannot detect type IV of clones.

Mostaeen et al. [P67] have used five clone detection tools to identify the clone instance in 12 software Java projects. An expert developer then evaluated the detected instances to recognize false positives and labeled clone distance. A machine learning model has been applied to classify the clone and non-clone samples. This data is then given as input to the artificial neural network (ANN), which learns to classify clones. This method can identify type I, II, and III clones, and the precision of this method on different Java projects is between 89.5% and 100%. The recall is between 67.6% and 96.7%. A similar method has been proposed by Li et al. [P59] to classify clone and non-clone instances. A Siamese neural network (SNN) [124] has been used by Patel and Roopak [P129] to detect type III and type IV code clones. However, their evaluations show higher precision for non-Siamese architecture. TPCaps [P132] uses a capsule network model [125] to identify clone pairs based on the features extracted from the program dependence graph (PDG). Saini et al. [P72] have proposed a learning-based method to detect clones in the twilight zone (clones between type III and Type IV) at the method level. First, the programs' methods are categorized concerning the number of tokens. A set of metrics is extracted for each method, and similar clones are identified based on vectors of metrics associated with each method. If the metrics of each method are exactly equal, it recognizes them as clones. However, if the values of the metrics are not equal, the clone pair is given as input to a Siamese deep neural network (SDNN) [124], which has already been trained to recognize clones using training data and determines whether this input is cloned. The method has an average recall of 69.78% and a precision of 89.5%.

Islam et al. [P124] have proposed a learning-based method for predicting buggy code clones. Six evolutionary features of code clones consist of change count, last commit of change, clone type, coupling, SPCP (Similarity Preserving Change

Pattern) extracted by detecting their genealogies and change patterns from thousands of revisions of subject systems and used those features to train a support vector machine for predicting buggy code clones. Their method can predict buggy code clones on different Java and C projects with overall 76% precision, 71% recall, and 73% F1.

An emerging application of code similarity measurement considered mainly by the learning-based approach is code recommendation, in which a code snipped is suggested to developers according to their needs [P85], [P133]. For instance, a model suggests the name for new software entities, *e.g.*, new methods, in the code based on similar entities at a benchmark [26], [126], [127]. Code2vec [P85] and Code2Seq [24] have used a code embedding approach to convert every method in the program to a fixed-length vector. A similar method body is then found by computing the vector similarity in a large corpus of programs' methods. Code2Vec [P85] and Code2Seq [24] use different paths between AST leaves of the methods as input to a deep neural network to learn the method embedding. Both approaches suffer from high computational cost and relatively weak performance in suggesting names for methods with complex AST have remained.

Tufano et al. [P65] have composed four representations of the code, including identifiers, ASTs, bytecode, and CFGs, to enhance the feature space of the learning model. The authors have concluded that using bytecode and CFG in the combined model improves the precision of code clone detection. SEED [P130] combines the data flow and control flow to form the semantic graph based on intermediate representation while focusing on API calls and operator tokens. The graph match network (GMN) [128] has been used to extract the feature vector from the semantic graph. The code fragments are required to be compilable for use in SEED.

Guo et al. [129] have proposed a pre-trained model that considers the inherent structure of code to provide models for various code-related tasks, including source code search, code translation, code refinement, and clone detection. In the pre-training stage, the authors have used data flow, a semantic structure of code that encodes the relation of "where-the-value-comes-from" between variables, instead of taking the syntactic structure of code like the AST. Their model is based on the transformer used in BERT's model [130]. About 2.3 million functions from six programming languages, including Java, Python, PHP, Go, Ruby, and JavaScript, were used for training. Their approach has achieved a precision of 94.8% and a recall of 95.2% in identifying code clones, indicating a good performance for code-related tasks. Tao et al. [P116] have exploited the CodeBERT model to detect cross-language code clones.

Chochlov et al. [P122] have used CodeBERT deep neural network to embed each code fragment in a fixed-length feature vector for detecting clones of types III and IV. The authors have then applied an efficient approximate $k$-nearest neighbor ($k$-NN) algorithm to avoid $O(n^2)$ comparisons between the embeddings for each of the $n$ code fragments. Their results show a lower execution time compared to SourcererCC [P47], [P53] and CCFinder [P5]. However, their approach requires systems with graphical processor units (GPUs).

It is unclear whether models like BERT and its variants provide the best pre-training when applied to other modalities, such as source code. Roziere et al. [131] have used pre-trained models based on masked language modeling (MLM) that leverages the structural aspect of programming languages [130] to recover the original version of obfuscated source code. The authors have shown that their pre-trained model considering program structural aspects significantly outperforms existing approaches on multiple downstream tasks. It provides relative improvements of up to 13% in unsupervised code translation and 24% in natural language code search. However, such a pre-trained model is often highly complex, and it is difficult to interpret their internal decision-making process. Changing the programming styles and renaming many parts of code could be adversarially used to destroy the results of such models. Sheneamer et al. [P112] have concluded that effective feature generation is more important than the type of classifier for machine learning code clone detection.

### 4.4.7   Image-based methods

Image-based code similarity measurement methods convert code fragments to the image and then use image processing techniques to find the similarity between code fragments [132]. Inspired by visual clone detection, Wang and Liu [P77] have proposed an image-based method for clone detection in which the preprocessed source codes are converted to images, and the Jaccard similarity between the images is computed to find similar source codes. The source code preprocessing steps mainly include removing comments and white space and highlighting keywords, data types, identifiers, and literals in the code. This approach cannot detect type IV clones or semantically similar codes, since it mainly relies on the appearance of the code snippets.

### 4.4.8   Test-based methods

Static source code analysis, used by most code similarity measurement approaches, is unsuitable for identifying semantically equivalent methods or type IV clones. Test-based methods use dynamic analysis in which the programs are executed with sample inputs or test suites, and their runtime data are collected. These data can then be used to detect semantically equivalent codes. Li et al. [P66] have proposed a test-based clone detection approach to identify type IV of code clones. A test suite is automatically generated for each pair of methods using the EvoSuite test data generation tool [133]. If two methods generate the same output on each of the generated test cases, they are considered semantically equivalent methods. The authors could identify clone methods of type IV in the Java development kit (JDK). However, generating and executing test cases for different methods is computationally expensive. Moreover, test-based methods cannot identify equivalent fragments within the body of methods. A similar method has been proposed by Su et al. [P50]. They have used use existing workloads to execute programs and then measure functional similarities between programs based on their inputs and outputs.

Leone and Takada [P123] proposed a test-based approach for type IV clones or semantic clone detection that overcomes the limitations related to objects in Java. The authors have used EvoSuite [133] to automatically create the tests with correct instantiations of all the needed classes in Java programs. The output of the test code for each method is compared using the DeepHash function to obtain numerical values for objects, considering the values of each instance variable. Their approach handles only methods which have a return value to compare. In addition, test data generation with EvoSuite is a time and resource-consuming process.

### 4.4.9   Hybrid methods

Hybrid methods combine two or more base methods to address the challenge of individual methods. Table 3 shows the different combinations of methods proposed by our primary studies and their application domains. It is observed that text-based and token-based methods have been mostly combined with graph-based and tree-based methods. The former methods are efficient but less effective, while the latter methods are effective but less efficient. Therefore, their combination provides efficiency and effectiveness for code similarity measurement. Token-based and text-based methods have also been combined to achieve high robustness while preserving the efficiency of text-based in finding types I, II, and III of clones [P20], [P29], [P55]. Finally, tree-based and learning-based methods have been mostly combined to appropriately detect near-missed and semantic clones [P80], [P81], [P82], [P83], [P97], [P125], [P126], [P134], [135].

Token-based and tree-based methods have been combined by Chilowicz et al. [P36] to detect clone types I, II, and III at the method level. The cloned instances are first selected based on their tokens' sequences. In the next step, the call graph for similar methods is computed and compared to filter out the false-positive instances specified in the first step. A similar approach has been proposed by Guo et al. [P131]. A combination of token-based and tree-based methods has been proposed by Vislavski et al. [P76] to detect cross-languages clones in five different programming languages, including Java, C, JavaScript, Modula-2, and Scheme. Two sequences of tokens extracted from pair of code fragments are compared together using a variant of the longest common sequence algorithm. However, the extracted sequence from ASTs can be too long, which makes comparison inefficient. Wang et al. [P136] have used AST to extract the required information such as type and context for building semantic tokens. The authors have applied n-gram to model each token based on $n$ previous variables in the AST. An F1 score of 90% has been reported for the proposed approach. However, building AST and n-grams for large codebase are time and resource computing.

Metric-based and graph-based have been similarly combined by Wang et al. [P40]. The candidate clones are chosen with a metric-based method, and then the results are filtered by comparing the PDGs of clone instances. Metric-based and graph-based have also been used together by Roopam et al. [5] to reduce the high cost of creating PDG for the entire program.

Kodhai et al. [P39] have combined the metric-based and text-based methods to detect clone instances at the method level. Candidate clones are detected by a metric-based, and then the results are filtered by a text-based approach to reduce false-positive samples. Higo et al. [P38] have detected clone instances at the method level by measuring the Jaccard similarity between the two sequences of tokens. The selected clones' names are then compared to identify the final samples.

Tekchandani et al. [P54] have proposed an algorithm based on reaching definition and liveness analysis to find semantically similar codes or clones' type IV. As a new application, their approach targeted programs developed in IoT domains. They first used a token-based method to select candidate clones and then compared the CFG of the selected candidates to determine the final groups of clone instances. They have leveraged both the data and control flow analysis to find irrelevant statements in each pair of code snippets. The authors conclude that the number of clone pairs increases as the number of dead statements increases. However, the proposed approach has not been evaluated empirically regarding the precision and recall measures.

Token-based and graph-based methods have also been combined by Yang et al. [P90]. However, they have used AST for graph comparison. Dong et al. [P104] have used a hybrid method to detect plagiarism in students' homework. Their approach first recognizes the style of the code by comparing the given code with a database containing different code styles. The ASTs of the source code with the same styles are created and compared to determine the final cloned instances. Thaller et al. [P127] have used token-based and test-based to identify semantic clones. The types of input and output parameters are compared in the static analysis phase using tokenized code fragments. Thereafter, the input and output value of the executable fragment (*e.g.*, function) is compared to decide the similarity between each pair of programs. Their proposed tool, SCD-PSM, cannot detect types 2 and 3 of code clones since textual similarities represent a different problem set. Tree-based and graph-based have been combined by Fang et al. [P99] to extract both the syntax and semantic features required for code similarity measurement, specifically detecting functional code clones.

Recently, learning-based, tree-based, and graph-based methods have been combined by many researchers to improve the feature space of the machine learning models for clone detection [P48], [P80], [P81], [P82], [P94], [P97], [P106], [P107], [P109], [P134]. Sheneamer and Kalita [P48] have manually extracted features from AST and PDG to train a classifier detecting code clones with type 4. Zeng et al. [P80] have automated the feature extraction from AST using deep autoencoder models [134]. Hu et al. [P125] have proposed TreeCen to detect semantic clones while satisfying scalability. First, AST is extracted from the given method and transformed into a simple graph representation called the tree graph, according to the node type. Thereafter, six types of centrality measures are extracted from each tree graph forming the feature vector of a machine learning model. The feature vector for a code pair, obtained by concatenating the two vectors, is labeled according to whether it is a clone or a non-clone pair. TreeCen [P125] only works at the method level. A similar method has been proposed by Wu et al. [P126] in which source code AST has transformed into a Markov chain instead of a tree graph.

Some hybrid-based clone detection approaches are not solely based on combining the aforementioned methods. They mainly rely on low-level code representations provided by programming language compilers [P20], [P28], [P96], [P117]. Selim et al. [P20] have proposed a hybrid clone detection technique to detect type III clones by combining source code and intermediate code processing. Their approach complements text-based and token-based clone detectors to detect type III clones. The limited number of operations in the intermediate representation decreases the dissimilarity in cloned code segments and leads to locating clone instances with complex variations.

Raheja et al. [P28] have computed metrics from the byte code representation to improve the performance of token-based clone detection. While these approaches increase the recall of clone detection compared to source-based methods, it requires compilation and decompilation of the program source codes to the intermediate codes. Schäfer et al. [P111] have provided a stubbing tool that effectively compiles Java source codes with missing dependencies into the Bytecodes. Their tool can be used for compiling code fragments whose dependencies are not presented within the dataset, such as the methods available in the BigCloneBench [135]. The authors have reported that their tool makes 95% of all Java files compilable in BigCloneBench which facilitates finding code clones with type I, II, and III.

Table 3. Various combinations of code similarity measurement and clone detection approaches.

| Base methods | Application domain | Main references |
|---|---|---|
| Token-based + Text-based | Clone detection | [P20], [P29], [P55] |
| Token-based + Graph-based | Clone detection | [P27], [P54], [P69], [P90] |
| Token-based + Test-based | Clone detection | [P127] |
| Metric-based + Token-based | Clone detection | [P28], [P62] |
| Metric-based + Graph-based | Clone detection | [P40] |
| Metric-based + Text-based | Clone detection | [P39] |
| Graph-based + Tree-based | Clone detection | [P48], [P99] |
| Text-based + Tree-based | Clone detection | [P14], [P70], [P111] |
| Token-based + Tree-based | Clone detection | [P36], [P76], [P131], [P136] |
| Graph-based + Learning-based | Clone detection | [P73], [P94] |
| Tree-based + Learning-based | Clone detection | [P80], [P81], [P82], [P83], [P97], [P125], [P126], [P134], [135] |
| Tree-based + Learning-based | Plagiarism detection | [P104] |
| Tree-based + Learning-based | Code recommendation | [P107] |
| Text-based + Graph-based | Clone detection | [P52] |
| Text-based + Tree-based + Graph-based | Clone detection | [P106] |
| Token-based + Metric-based + Learning-based | Clone detection | [P72] |
| Text-based + Tree-based + Graph-based + Learning-based | Clone detection | [P109] |

## 4.5    Code similarity measurement tools and languages

We could find 80 academic software tools proposed by primary studies in our SLR. More precisely, 80 of 136 primary studies have indicated a tool supporting their proposed method. However, not all of these tools are publicly available. Table 4 shows the tool names along with their download links, methods, supporting languages, and development languages. The tools whose download links and supporting languages are not mentioned in the literature have not been reported in Table 4. The complete list of 80 tools proposed in primary studies is available in our online Excel appendix [80]. Only 33 out of 80 (about 41%) articles, pointed to a tool name, have publicized their tools, indicating the lack of tool-supported and reliable publications in this field. We could not find any working link to the remaining 47 academic tools.

It is worth noting that there are code similarity measurement and clone detection tools not discussed in the research papers but we have listed them in Table 4 to provide a complete reference for researchers and practitioners. The CPD tool is a submodule in the PMD source code analyzer [86] that finds duplicated code in more than 20 programming languages, including but not limited to C, C++, C#, Go, Groovy, Java, JavaScript, and Matlab. Simian [89] is another tool that identifies duplication in Java, C#, C, C++, COBOL, Ruby, JSP, ASP, HTML, XML, Visual Basic, Groovy source code, and even plain text files. iClones [87] extracts clone evolution data from a program's history using the token-based technique. iClones has been used by Ehsan et al. [P135] to extract clone instances from programs' history on the public GitHub repository. The authors have used machine learning to rank the retrieved clones based on their faultiness and enhance maintenance activities. MOSS [136] and AutoMOSS [88] are text-based source code similarity measurements dedicated to plagiarism detection.

Table 4. Existing code similarity measurement tools sorted alphabetically based on the tool name.

| Tool name | Link to the software tool | Method | Application | Supporting language(s) | Programming language(s) |
|---|---|---|---|---|---|
| Amain | *https://github.com/CGCL-codes/Amain* | Tree-based + Learning-based | Clone detection (Type IV) | Java | Python |
| AST+ | Not mentioned | Tree-based | Defect prediction | C | C |
| ASTNN | *https://github.com/zhangj111/astnn* | Tree-based + Learning-based | Clone detection (types I, II, III-IV) | Java, C | Python |
| AutoMOSS | *https://github.com/automoss/automoss* | Text-based | Plagiarism detection | Many-languages | Python |
| CCAligner | *https://github.com/PCWcn/CCAligner/tree/f27622d6f1500 536c45862c4d49bd5f5d6802ace* | Token-based | Clone detection | Java | C++ |
| CCfinderx | *https://github.com/gpoo/ccfinderx* | Token-based | Clone detection (types I, III) | Java | C++, Java, Python, C |
| CCSharp | *https://github.com/pquiring/CCSharp* | Graph-based | Clone detection (all types) | C# | C# and C++ |
| CCStokener | *https://github.com/CCStokener/CCStokener* | Token-based + tree-based | Clone detection (types III) | C, Java | Python |
| Clone Manager | *https://www.scied.com/pr_cmbas.htm* | Token-based + tree-based | Clone detection (all types) | C, Java | Not mentioned |
| Clone Miner | Expired link | Learning-based | Clone detection | Java, C++, Perl, VB | Not mentioned |
| Clone TM | Not mentioned or expired link | Text-bases | Clone detection (types I, III) | Java, C++, C# | Not mentioned |
| CloneCognition | *https://github.com/pseudoPixels/CloneCognition* | Learning-based | Clone detection | Java | Javascript-Python |
| CLONE-HUNTRESS | Expired link | Tree-based | Clone detection (types I, II, III) | Java | Java |
| Clonewise | Not mentioned | Learning-based | Clone detection | C | Not mentioned |
| CP-Miner | Not mentioned or expired link | Token-based | Clone detection | C, C++ | Not mentioned |
| CrolSim | *https://github.com/Kawser-nerd/CroLSim* | Learning-based | Similarity detection | Java, C++, C#, Python | Python |
| Deckard | *https://github.com/skyhover/Deckard* | Tree-based | Clone detection (types I, II, III) | Java | C, C++, Python |
| DeepSim | *https://github.com/parasol-aser/deepsim* | Graph-based + learning-based | Similarity detection | Java | Java, Python |
| Deimos | Not mentioned or expired | Token-based | Plagiarism detection | LISP, Pascal, C | Not mentioned |

| Tool name | Link to the software tool | Method | Application | Supporting language(s) | Programming language(s) |
|---|---|---|---|---|---|
| DyCLINK | *https://github.com/Programming-Systems-Lab/dyclink* | Graph-bases | Clone detection (types I, II, III) | Java | Java |
| FCDetector | *https://github.com/shiyy123/FCDetector* | Tree-based | Clone detection | C, C++ | Java, Matlab |
| gCad | Not mentioned or expired link | Text-based | Clone detection | C, C#, Java | Java |
| Groupdroid | Not mentioned or expired link | Graph-based | Malware detection | Java | Not mentioned |
| HitoshiIO | *https://github.com/Programming-Systems-Lab/ioclones* | Text-based | Clone detection | Java | Java |
| iClones | *http://www.softwareclones.org/downloads.php* | Token-based | Clone detection (types I, II, III) | Java, C | Java |
| IDCCD | *https://github.com/MRHMisu/Interface-Driven-Code-Clone-Detection* | Token-based | Clone detection (types I, II, III) | Java | Java |
| JPlag | *https://github.com/jplag/jplag* | Token-based | Plagiarism detection | Java, Scheme, C, C++ | Java |
| LICCA | *https://github.com/tvislavski/licca* | Token-based + tree-based | Clone detection (cross-language) | Java, JavaScript, C, Modula-2, and Scheme | Java |
| LSC-Miner | Not mentioned or expired link | Token-based | Clone detection (cross-language) | C, C++, Java, VB6, VB.Net, C# | C, C++, Java |
| MLClone Validation Framework | *https://github.com/pseudoPixels/ML_CloneValidationFramework* | Learning-based | Clone detection | Java | Not mentioned |
| MOSS | *https://theory.stanford.edu/~aiken/moss* | Text-based | Plagiarism detection | Many-languages | Not mentioned |
| Mutation Injection Framework | *https://github.com/jeffsvajlenko/MutationInjectionFramework* | Tree-based | Clone detection (all types) | Java, C, C# | Not mentioned |
| NiCad | *https://github.com/bumper-app/nicad* | Text-based | Clone detection (types I, II, III) | Java | Java, C, HTML |
| Oreo | *https://github.com/Mondego/oreo* | Token-based + metric-based + learning-based | Clone detection (all types) | Java | Java, Python |
| PlaGate | Not mentioned or expired link | Text-based | Plagiarism detection | Java | Not mentioned |
| PMD/CPD | *https://github.com/pmd/pmd* | Token-based | Clone detection (types I, II, III) | Many-languages | Java |
| SEED | *https://github.com/ZhipengXue97/SEED* | Learning-Based | Clone detection (type IV) | C, Java | Not mentioned |
| SENSA | *https://github.com/m-zakeri/SENSA* | Metric-based | Code recommendation | Java | Python |
| Siamese | *https://github.com/UCL-CREST/Siamese* | Token-based | Clone detection (types I, II, III) | Java | Java, Python |
| Simian | *https://simian.quandarypeak.com/* | Token-based | Clone detection (types I, II, III) | Many-languages | Java, .NET |
| SourcererCC | *https://github.com/Mondego/SourcererCC* | Token-based | Clone detection (types I, II, III) | Java, C++, JavaScript, Python | Java, Python |
| Stubber | *https://github.com/andre-schaefer-94/Stubber* | Text-based + Tree-based | Clone detection (types I, II, III) | Java | Java |
| TECCD | *https://github.com/YangLin-George/TECCD* | Tree-based | Clone detection (types I, II, III) | Java | Java |
| Tisem | *https://github.com/3120150428/Tisem* | Test-based | Clone detection | Java | Not mentioned |
| TPCaps | Not mentioned | Learning-based | Clone detection (all types) | Java | Not mentioned |
| TreeCen | *https://github.com/CGCL-codes/TreeCen* | Tree-based + learning-based | Clone detection (Type IV) | Java | Python |

As shown in Figure 8, most available tools use hybrid, learning-based, tree-based, or token-based methods in their backend. The reason is the high efficiency of these methods and compatibility with most programming languages. On the other hand, only one tool works with the test-based approach indicating the difficulty and novelty of research in this direction.

Figure 9 shows the frequency of programming languages supported by available source code similarity measurement and clone detection tools. The languages supported by only one tool are shown separately on the right-hand side of the pie chart diagram. About 7% of the supported programming languages are supported by only one tool. The top five frequently supported languages are Java, C, C++, C#, and Python. Nearly 44% of available tools support the source codes written in Java, and 33% support C or C++ codes, while other popular programming languages like PHP, Swift, and Scala have no supporting tool. As a result, there is a vast opportunity to develop multi-paradigm and multi-language code similarity measurement and clone detection tools.

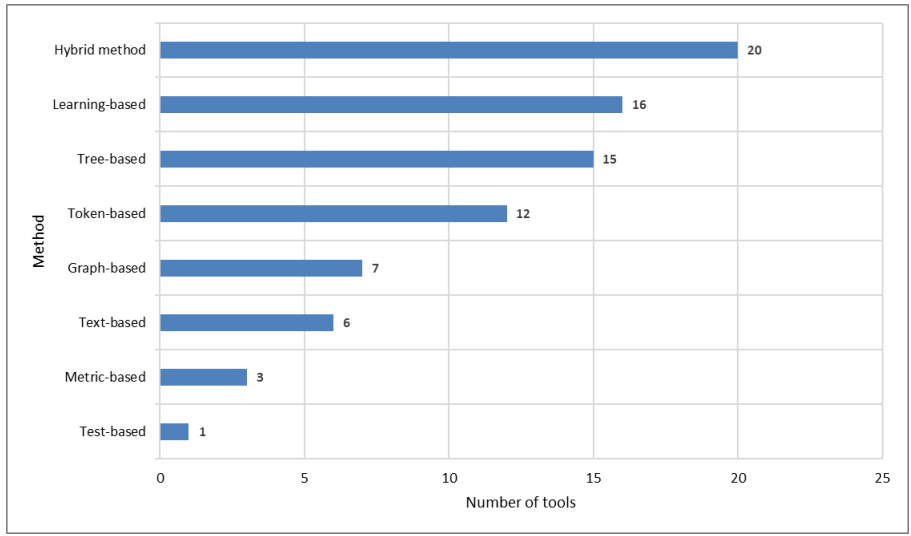

Figure 8. Number of academic tools built based on each method.

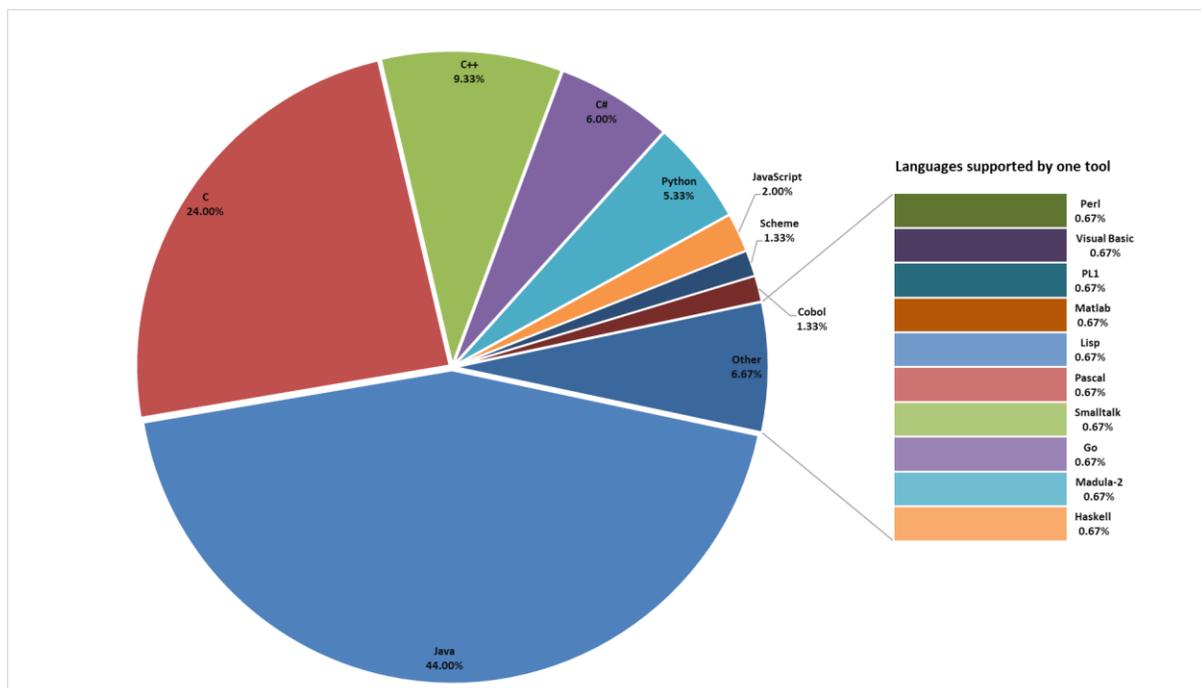

Figure 9. Frequency of programming languages supported by code similarity measurement tools.

## 4.6    Code similarity benchmark and datasets

Dataset plays an essential role in building and validating code similarity measurement tools. The investigation of primary studies demonstrates that there are 12 datasets concerning source code similarity and clone detection research. Table 5 lists the available code similarity datasets and some additional information download links, supporting languages, size, and their main reference study. Only 68 out of 136 (50%) studies have used these datasets. The remaining 50% of the primary studies have mainly used a limited set of open-source projects, indicating the lack of a reliable evaluation for almost half of the proposed approaches. Our investigation shows that over 70 different open-source projects have been used in these articles. Most studies have only reported using GitHub projects based on their stars with no details, making it difficult to compare the different approaches fairly. Figure 10 shows the word cloud plot of the projects used to create and evaluate code similarity measurement tools based on the information reported in the primary studies. It is observed that the primary studies have mainly used Apache Ant, ANTLR, JDK, and PostgreSQL projects to build and validate their approaches.

Bellon et al. [62] have created a reference corpus for clone codes of types I, II, and III in eight software suites (four C programs and four Java programs). Bellon's benchmark represents locational information of code clones with the file name, start line number, and end line number. The benchmark dataset does not contain the locational data of gapped lines. Hence, it is not possible to evaluate some type III clones correctly [137]. Murakami et al. [137] have added the locational information of the gapped lines to Bellon's dataset. However, we observed that the public link to this dataset was obsolete. Experiments by Charpentier et al. [138] show that a significant amount of the reference clones are debatable, and this phenomenon can introduce noise in results obtained using Bellon's benchmark.

BigCloneBench (BCB) [92], [135], [139] is a clone detection benchmark of known clones in the IJaDataset source repository. The current version of the benchmark, IJaDataset (with modifications), and tools for measuring clone detection recall are publicly available on a GitHub repository, shown in Table 5. The source code files in the BCB dataset have been crawled from various Java open-source projects. To the best of our knowledge, BCB is the largest dataset for clone detection and code similarity measurement. It consists of projects from 25000 systems and covers ten functionalities, including 6000000 true clone pairs and 260000 false clone pairs annotated by the domain experts. Krinke and Ragkhitwetsagul [140] have criticized the way that BCB [92], [135], [139] has been constructed and stated that it is problematic to be used as ground truth for learning code similarity measurement, specifically with machine learning techniques. For example, BCB contains true but unlabeled clone pairs for different functionalities. In addition, the BCB dataset is imbalanced and biased. The authors have not proposed any solutions to fix the existing drawbacks of BCB.

The CDLH work [141] presents a filtered version of BCB in which the authors have discarded code fragments without any true or false tags for the clone pairs. As a result of such filtering, the CDLH dataset includes 9134 code fragments. However, the authors have not published the filtered dataset publicly. Schäfer et al. [142] have investigated the role of ground truth quality in learning-based clone detection tasks. The authors have concluded that the CDLH dataset [141] is not suitable for evaluating learning-based clone detection tools. Both the BCB and CDLH datasets suffer from two main problems: (1) Training, validation, and evaluation sets are not strictly separated. (2) There is no ground truth on negative samples, *i.e.*, what is not a code clone.

Lu et al. [143] have recently introduced a machine learning benchmark dataset, CodeXGlue, for code understanding and generation. CodeXGlue includes a collection of ten tasks across 14 different datasets, of which two datasets belong to the code clone detection tasks. CodeXGlue uses 9126 samples of the CDLH dataset and all samples of the OJClone/POJ dataset [144].

The OJClone/POJ (POJ-104) dataset [144] contains 104 programming problems and their solutions' source codes in C/C++, submitted by 500 students for each problem. A pedagogical programming open judge (POJ) system automatically examines the validity of submitted source code for specific problems by running the code with test inputs. Different source codes solving the same programming problem are considered clone instances. CF-500 [145]contains more than 23,000 code snippets written in C, covering 500 problems from the open judging platform Codeforces [95]. Each two code snippet can construct a code pair. Therefore, the number of code pairs in a dataset can be higher than 10,000,000 if using all possible combinations. Code4Bench [91] is a similar dataset created on top of the source code submitted to the Codeforces programming competition platform [95]. Code4Bench has mainly been used for fault prediction tasks. The SOCO dataset proposed by Flores et al. [146] focused on detecting re-used source codes in C/C++ and Java programming languages. It includes 427 pairs of clone codes and 66628 non-clone pairs.

To increase the size and variety of the code similarity datasets, Zhao and Huang [147] have collected a set of 1669 projects from 12 competition problems in the Google CodeJam competition [94]. The code snippets submitted for each problem in this dataset have similar functionality, and Google has validated them. Therefore, they can be used for building and evaluating tools to detect functional similarity (type IV of clones). There are other online programming competition platforms such as AtCoder [148] whose data have been used by Perez and Chiba [149]. However, the most frequently used platforms are Google CodeJam competition [94] and Codeforces [95].

The GPLAG dataset has been created by Liu et al. [16] to evaluate plagiarism detection with PDG-based methods. The authors have selected five procedures from an open-source Linux program, "*join,*" and applied a set of plagiarism operators [16] to generate a modified version of the program. The dataset is not publicly available. However, it can be reconstructed according to the full descriptions provided by the authors. The Malware dataset [20] has been proposed to evaluate the malware detection task. It includes original malicious codes, variants of the malware, and harmless codes written in Visual Basic language. The GPLAG and Malware are small datasets that cannot be used in learning-based approaches. Overall, it observed that datasets regarding many applications of code similarity measurement, such as code recommendation, code prediction, malware, and vulnerability detection are limited.

Table 5. Dataset proposed for code similarity measurement and clone detection tasks.

| Dataset | Download link | Programming language | Samples (Size) | Ref(s). |
|---|---|---|---|---|
| BigCloneBench (BCB) | *https://github.com/clonebench/BigCloneBench* | Java | 6260000 code pairs (~2.3 GB) | [92], [135], [139] |
| CDLH | Not available publicly | Java | 9134 code snippets | [141] |
| Bellon's benchmark | *http://www.softwareclones.org/research-data.php* | C and Java | 93 clone pairs | [62] |
| OJClone/POJ (POJ-104) | *http://programming.grids.cn* | C, C++ | 52000 code snippets | [144] |
| CF-500 | *https://github.com/ZhipengXue97/SEED* | C | 2300 code snippets | [145] |
| Code4Bench | *https://zenodo.org/record/2582968* | C, C++, Java, Python, and Kotlin | (~0.63GB) | [91] |
| SOCO | *https://pan.webis.de/fire14/pan14-web/index.html* | C, C++, and Java | 67055 code pairs | [146] |
| Google CodeJam | *https://code.google.com/codejam/contests.html* | Java | — | [147] |
| CodeXGlue | *https://github.com/microsoft/CodeXGLUE* | Java, C, and C++ | 61126 code snippets | [143] |
| GPLAG | Not available publicly | Java, C | — | [16], [150] |
| Malware | Expired | Visual Basic | — | [20] |
| ANSI-C | Not available publicly | C | — | [151] |

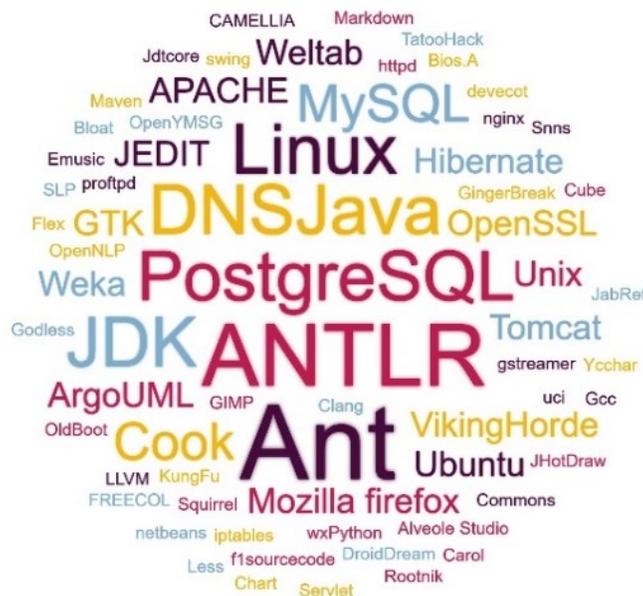

Figure 10. Word cloud of projects used to create and evaluate code similarity measurement tools.

## 4.7    Comparison of code similarity measurement techniques

In this section, we compare existing techniques and applications in code similarity regarding their performance in different application domains. The code similarity measurement techniques are evaluated through the three factors of *effectiveness*, *efficiency*, and *scalability*. The effectiveness is typically measured with precision, recall, F1, and accuracy metrics and determines how the approach performs well on different data samples, while the efficiency is related to the speed and computation cost of the proposed technique. Finally, scalability is concerned with the factors affecting the execution of the algorithms on large codebases.

Unfortunately, not all primary studies reported the results for the discussed evaluation metrics, which means further empirical studies are required in the area of code similarity evaluation. Due to the different benchmarks, datasets, and applications used by primary studies, it is difficult to perform a precise and fair quantitative analysis of the existing methods. We, therefore, compared the results reported in each study using qualitative terms (high, moderate, and low) for each evaluation metric according to the result reported in primary studies. Table 6 summarizes the result of a qualitative comparison of the discussed methods. We have mentioned supporting clone types, effectiveness, efficiency, and scalability for each technique. We also have listed the essential pros and cons of the approaches. The hybrid approaches, which combine more than two based approaches or are suggested by only one study, have not been mentioned in Table 6 because we could not find sufficient evaluation results due to the limited number of papers with such methods.

The F1 values reported by the relevant primary studies with similar systems were averaged to determine the overall effectiveness of a method. The "high" term in the third column of Table 6 denotes average effectiveness greater than 0.85, the "moderate" term used for average effectiveness between 0.70 and 0.85, and the effectiveness lower than 0.70 means "low" effectiveness. The efficiency and scalability have been evaluated according to the execution time, required resources, and complexity of the proposed algorithms discussed in primary studies.

It is observed that none of the similarity measurement techniques can detect all types of clones with an effective, efficient, and scalable mechanism. Different types of clone detection might be required based on the application domain. As mentioned in Table 6, all proposed techniques, even the hybrid ones, contain specific advantages and disadvantages when they come into practice. For instance, metric-based methods have relatively high recall and scalability, while their performance is highly sensitive to the threshold applied to metrics values or vector similarities. It also should be noted that we could not find any publicly available tools that work with this method. Token-based code similarity measurement tools, including IDCCD [6], Siamese [152], SourcererCC [153], SourcererCC-I [154], CPPCD [155], SCDetector [156] CCFinderX [2], CCAligner [157], and iClones [87] are suitable for clone types I, II, and III. They expose relatively high effectiveness, efficiency, and scalability compared to other methods. However, they cannot achieve good accuracy in clones of type IV. Graph-based code similarity measurement tools, including CloneDetective [158], CBCD [103], CodeBlast [159], SPAPE [160], CCSharp [83], CCGraph [161], and Groupdroid [18], HOLMES [162], and SEED [163] work well on type IV clones. However, they are not efficient and scalable [164]. In addition, graph-based tools are vulnerable to techniques such as code obfuscation which is typically applied to malicious codes. Learning-based tools such as Clone Miner [165], CCLearner [166], CrolSim [60], DeepClone [167], and CCEyes [168] are accurate and scalable, but they often require huge and clean datasets for each application domain to work properly. Recent learning-based tools, including Code2Vec [23] and Code2Seq [24], are suitable for code recommendation tasks. However, they only work at method-level similarity measurements. It is possible to extend them for working on different applications and abstraction levels. The comparisons imply that the best approach must be selected according to the user's preference, application domain, computational resource, and non-functional requirements.

Table 6. A qualitative comparison of code similarity measurement methods.

| Approach | Supporting clone types | Effectiveness | Efficiency | Scalability | Advantages | Disadvantages |
|---|---|---|---|---|---|---|
| Text-based | I, II | High | High | Low | Straightforward implementation | • Requires heavy preprocessing<br>• Fails to identify types III and IV clones |
| Token-based | I, II, III | High | High | High | High ability and high compatibility | • Fails to identify type IV clones<br>• Needs lexer transformation rules |

| Approach | Supporting clone types | Effectiveness | Efficiency | Scalability | Advantages | Disadvantages |
|---|---|---|---|---|---|---|
| Tree-based | I, II, III | High | Low | Moderate | High accuracy for clones with types II and III | • Fails to identify type IV clones<br>• Constructing AST is difficult and complex |
| Graph-based | I, II, III, IV | High | Low | Low | • Identifies all types of clones<br>• Considers both the structural and semantic information | • Graph matching is non-polynomial<br>• Building PDG and CFG are difficult and complex |
| Metric-based | I, II, III, IV | High | Moderate | High | • Identifies all types of clones<br>• Language independent<br>• Straightforward similarity measurement | • Very sensitive to metric thresholds<br>• Low recall |
| Learning-based | I, II, III, IV | Moderate | Low | High | • Identifies all types of clones<br>• Fully automated approaches<br>• High stability and flexibility | • Requires large and clean datasets |
| Token-based + Text-based | I, II, III | High | Moderate | Moderate | Straightforward implementation | Fails to identify type IV clones |
| Token-based + Graph-based | I, II, III, IV | High | Low | Moderate | Identifies all types of clones | Constructing PDG is difficult and complex |
| Metric-based + Token-based | I, II, III, IV | High | Moderate | High | • Identifies all types of clones<br>• Straightforward implementation | Requires lexer transformation rules |
| Metric-based + Graph-based | I, II, III, IV | High | Moderate | Moderate | • Identifies all types of clones<br>• More efficient compared to the graph-based method<br>• Suitable for source code visualization and visual clone detection | • Complicate implementation<br>• Very sensitive to metric thresholds |
| Metric-based + Text-based | I, II, III, IV | High | Moderate | High | • Identifies all types of clones<br>• More efficient compared to the graph-based method<br>• Low false-positive instances | • Requires heavy preprocessing<br>• Very sensitive to metric thresholds |
| Graph-based + Tree-based | III, IV | Moderate | Low | Low | Detects semantic clones | Cannot identify all types of clones |
| Text-based + graph-based | I, II, III, IV | High | Moderate | Low | Identifies all types of clones | • Building PDG and CFG are difficult and complex |
| Text-based + Tree-based | I, II, III | High | Moderate | Low | Low false-positive instances | Cannot identify all types of clones |
| Graph-based + Learning-based | I, II, III, IV | High | Moderate | Low | Identifies all types of clones | • Building PDG and CFG are difficult and complex<br>• Requires large and clean datasets |
| Tree-based + Learning based | I, II, III | High | Moderate | Moderate | More efficient compared to the graph-based method | • Cannot identify all types of clones<br>• Requires large and clean datasets |
| Token-based + Tree-based | I, II, III | High | High | Moderate | The preprocessing is used to increase the detection speed | Fails to identify type IV clones |

# 5    Challenges and Opportunities

Although many advancements have been made in code similarity measurement and clone detection approaches during the last two decades, there are some challenges and many research opportunities in this area. We found and prioritized six significant challenges that have not been addressed by the primary studies in this systematic literature review.

**Challenge 1**: One trivial challenge is the lack of reliable datasets for building and evaluating new code similarity measurement tools regarding different dimensions, including size, entity types, severity levels, clone types, and programming language. The size concerns the number of labeled instances in the dataset for different entity types in different programming languages [143], [147]. Entity types concern the granularity of the abstraction levels, which can be used in similarity measurements, including the statement, code block, method, class, and package. Specifically, statements and code-block similarity enable the applications of code similarity measurement in new areas such as automatic fault localization [169]–[171], fault prediction [172], and program repair [173]. The optimum similarity measurement technique for each entity type most presumably differs, which should be studied and discussed in future works. The severity level aims to measure similarity with more than two values of similar and dissimilar codes. Using fuzzification in this context helps better decision-making in some problems, such as estimating code debts. A dataset supporting all types of cloned code would be the preferable choice concerning clone type over the current datasets. Finally, datasets containing more than one programming language help evaluate the generalization of the proposed approach and cross-language clone similarity measurement. It is worth noting that the role of the dataset is not limited to validation purposes [174]. Datasets are essential artifacts in creating new code similarity measurement tools, specifically learning-based ones [129], [141], [147], [167], [168]. Proposing automated and semi-automated methods to create large-scale yet reliable code similarity datasets adds considerable value to the generalizability of the results reported in the field. For example, Roy et al. and Cordy [93] have proposed an automated method to synthesize large numbers of known clones using the 14 mutation operators regarding the four types of clones in Java. However, such machine-generated clones may not be similar to the ones created by developers. Pyclone generates type I, II, and III code clones for Python in the same way [175]. Yu et al. [96] have proposed a more advanced framework based on program transformation to increase the data samples in the already existing datasets containing real-world data. They have concluded that their data augmentation framework improves the performance of code similarity measurement tasks. One promising approach to identify and label more natural instances of clone codes is to search large code hosting providers such as GitHub [176] to find similar code fragments based on the content of comments and issues in different repositories and then validate the results manually.

**Challenge 2**: Some code similarity measurement methods, especially metric-based, learning-based methods, and test-based, require serious attention from researchers and practitioners. They can be promising when appropriately combined with other existing techniques. Metric-based methods in the previous research have used a maximum of 21 metrics plus Halstead's metrics. However, the number of source code metrics reported in the software engineering literature for different programming paradigms is much more than nine metrics. For instance, a total of 190 different metrics have been reported for the object-oriented paradigm [76]. Metrics-based approaches can be used to represent source codes as fixed-length vectors to shape the inputs required by learning-based methods. Increasing the number of source code metrics and applying automatic feature selection techniques enhances learning-based methods' feature space, resulting in more accurate code similarity measurement models. Recent works by Zakeri-Nasrabadi and Parsa [27]–[29] show the possibility of enriching feature space using new source code metrics. The newly introduced metrics improve the performance of machine learning models predicting test effectiveness which can be considered as another application of code similarity measurement. A better source code representation results in a better similarity measurement [177]. Source code metrics provide advantages such as language independence, interpretability, and efficiency in code similarity measurement [26]. These advantages are not observed in the deep learning approaches automatically extracting features from the source codes [23], [24], [129].

**Challenge 3**: An important factor affecting the application of proposed code similarity tools is their *efficiency*, more precisely, the time required to measure code similarity. Unfortunately, no quantitative analysis has been reported for the efficiency of code similarity measurement tools in different application domains. The low efficiency of some approaches, such as graph-based and tree-based, may prevent using them on a large codebase. Systematic empirical evaluations of the existing code similarity measurement tools' scalability and efficiency are required to find the optimum hybrid method to maximize these aspects. Some approaches, such as metric-based and learning-based, can parallelize and scale on demand. However, this requires more computational cost, which may not be acceptable in all situations. Hosting code similarity measurement tools on cloud infrastructures facilitates resource allocation and scalability. Embedding cache mechanisms in the tools also helps to avoid repeating all calculations when they are not necessary. Finally, techniques such as transfer learning can be applied to create code similarity measurement models for new programming languages efficiently from the trained model on similar languages.

**Challenge 4**: While most primary studies have used a hybrid method to measure code similarity, publicly available tools mostly work with the token-based method. Indeed, the number of software tools that work with more complicated procedures than text-based and token-based ones is limited. Researchers have rarely used the existing library for basic tasks such as graph comparison [178], machine learning [179], static analysis [180], and metrics computations [181] to build reliable and reusable tools. In addition, the tools have been rarely integrated with IDEs, making them difficult to be used by software engineers. We recommend preparing future tools as plugins for IDEs with appropriate API and documentation. The applications that rely on the core capabilities of the code similarity tools may be categorized within these APIs and plugins to facilitate developers' use.

**Challenge 5**: Most proposed methods have focused on finding clone instances and measuring code similarity in object-oriented programs, mainly written in Java. However, the recent increase in the popularity of multi-paradigm languages such as Python and JavaScript requires approaches supporting these paradigms. For instance, metric-based methods may employ only source code metrics available in all programming paradigms. Researchers are encouraged to focus on developing tool-supported approaches for other programming languages than Java and C/C++. The main challenge is adapting an existing method to work with a new language for which no code similarity dataset is most likely available. As mentioned earlier, one solution is to apply transfer learning, in which the existing models are fine-tuned to work with a new domain using only a few data samples. Another possible solution is to create models based on source code metrics that are independent of the language structures.

**Challenge 6**: The final challenge we observed is the lack of using code similarity techniques and tools to assist the development phase in the software development lifecycle (SDLC). Current clone detection tools have been designed to be used in the software maintenance phase. It means that they work on legacy programs, not on developing new programs. The real-time measurement and notification of code similarity during software development help programmers avoid repeating code and reuse the available software component. As discussed in this paper, the application of code similarity measurement is not limited to clone detection. Other applications mentioned in this paper, mainly code recommendations, improve software development's agility.

We believe that code similarity measurement can be used as a fundamental component for all data-centric solutions in software engineering. The general theory of code similarity enables the automation of many laborious programming tasks, including but not limited to automatic code recommendation, defect prediction, smell detection, vulnerability finding, refactoring recommendation, and quality measurements. Although researchers developed automated approaches separately in each domain, the relationship between most of these applications and code similarity measurement as a common principle has not been well studied. Investigating the common aspects of the aforementioned applications with respect to code similarity measurement help to develop versatile tools that can perform multiple tasks efficiently.

Automatic code recommendation, *e.g.*, software entity naming and code summarization, has been recognized as a problematic software engineering task [126], [182]. Code similarity measurement can be best applied to detect cloned instances and recommend the name of detected clones to programmers. Alon et al. [23], [24] have recently proposed a hybrid tree-based and learning-based code similarity measurement method to predict program properties such as names and expression types according to existing cloned or similar methods. The challenge is to find the clone instance with a clean and proper name. Appropriate entity names for a developing program may be found in similar codes in another programming language. In such cases, cross-language clone detection approaches can be worthful.

Code similarity measurement and clone detection can also be used for identifying code smell as refactoring opportunities. A recent study by Aniche et al. [35] has used machine learning techniques to recommend refactoring based on code similarity. Integrated development environments (IDEs) are expected to be equipped with various code similarity measurements to make online suggestions about existing clones, refactoring, faults, plagiarism, and other software quality attributes. In Software 2.0 [183], where programming uses learned models, the syntax and functionality of all codes are expected to be very similar, and what makes differences are the input and output data. Therefore, a new definition of the concept of code similarity is required to adapt for programs written in Software 2.0 [183]. In addition, a powerful code recommendation tool can be designed and developed to make a super agile and automated software development methodology.

# 6   Threat to Validity

The results of this survey might be affected by the article selection bias, incompleteness of search, search engine problems [184], study distribution imbalance, and inaccuracy in data extraction and synthesis. We tried to mitigate each threat as much as possible during the review process. First, we followed the well-defined research protocol guidelines proposed by Kitchenham and Charters [75] to select relevant studies fairly and without bias. Second, we defined our search string such that all relevant terms and their possible combination were covered to overcome the threat of missing a study. We also performed snowballing [79] to find any relevant paper in the field, initially not found by our search string. Third, we considered a quality assessment before the final selection of primary studies to ensure that the papers for all key terms were retrieved correctly and made a significant contribution and validation in the field. Forth, we carefully compared the taxonomy used in our classification with the related survey in code similarity measurement and code clone detection. Finally, we asked three M.Sc. students and one Ph.D. student in software engineering to check the correctness and completeness of our classification according to the research questions we aimed to answer.

## 7    Conclusion

This systematic literature study performs automatic searches in four major electronic libraries to select the relevant code similarity measurement and clone detection studies. A total of 136 studies are selected and reviewed in detail from an initial set of over 10000 articles to answer five research questions about different aspects of this topic. This paper analyzes each primary study according to five dimensions: method, application, dataset, tool, and supporting programming language. Our survey aims to answer five research questions concerning the proposed classification of the primary studies in code similarity measurement. The short answers to our research questions and relevant finding based on the discussions in this review are as follows:

**RQ1 findings**: Our SLR reveals the existence of at least eight different basic methods, including text-based, token-based, tree-based, graph-based, metric-based, image-based, learning-based, and test-based, used for source code similarity measurement. Learning-based and test-based approaches have recently been applied in the context of source code similarity measurement and clone detection and have not been covered by previous surveys. Our SLR indicates that most studies (over 27%) have employed hybrid methods in which code snippets' textual and structural contents are compared to measure code similarity. However, only 41% of articles have supported their proposed approaches with a publicly available software tool.

**RQ2 findings**: Our findings indicate that the research on code similarity measurement primarily targets a direct application of clone and reuse detection. However, we found four other application domains: source code plagiarism detection, malware detection and vulnerability, code prediction, and code recommendation. Code similarity measurement can automate laborious activities in software engineering, such as code smell detection, refactoring suggestions, and fault prediction.

**RQ3 findings**: We extracted 80 software tools for measuring source code similarity and clones from the primary studies, of which 33 tools (*i.e.*, 41%) are publicly available. In total, the available tools support code similarity measurement and clone detection for 18 different programming languages. More than 77% of these tools support programs written in Java, C, and C++ languages, demonstrating the lack of code similarity measurement tools for other programming languages and paradigms.

**RQ4 findings**: At least 50% of the primary studies have mainly used a limited set of open-source projects, including Java projects on GitHub. We observed the use of 12 different datasets designed explicitly for code clone detection and source code similarity measurement. Only 68 out of 136 primary studies (50%) have used these datasets, and not all of the reported datasets are publicly available. We observed the lack of public, large, and quality source code similarity and clone datasets containing industrial and real-life software systems.

**RQ5 findings**: The performance of code similarity measurement studies is evaluated with different metrics in three dimensions effectiveness, efficiency, and scalability. Regarding the effectiveness of existing techniques, our meta-analysis shows an approximate mean of 86.3, 88.4, 86.5, and 82.5 percent, respectively, for the precision, recall, F1 score, and accuracy metrics. However, these results are based on the different datasets used to evaluate the existing tools and they are only moderately reliable. The performance of emerging applications based on code similarity, *e.g.*, code recommendation, is less than older applications. Further empirical evaluation of code similarity measurement and standard datasets are required in all application domains to determine the state-of-the-art.

**RQ6 findings**: Our SLR identifies and lists six remarkable challenges in the field with some potential solutions that can be considered as the future direction of research on source code similarity and clone detection. The lack of comprehensive, large, and reliable datasets, lack of attention to metric-based and learning-based methods, limited support of popular programming languages and new programming paradigms, lack of empirical analysis on the efficiency of scalability of different approaches, and considering the emerging applications based on code similarity measurement are the most critical challenges and opportunities in the field.

Research on code similarity measurement and its applications is growing and will continue to grow in the following years. Industrial support of proposed approaches with reliable tools is essential to reduce the high cost and time of developing quality software systems. At the same time, systematic literature reviews and empirical studies in the field are also necessary to integrate the results and provide helpful information to practitioners and researchers.

## Declarations

### Data Availability Statement

The datasets generated and analyzed during the current study are available on Zenodo, https://doi.org/10.5281/zenodo.7993619.

### Funding

This study has received no funding from any organization.

### Conflict of Interest

All of the authors declare that they have no conflict of interest.

### Ethical Approval

This article does not contain any studies with human participants or animals performed by any of the authors.

## Appendix A: Primary Studies

Table 7 shows the primary studies used in the SLR. The papers are sorted based on their publication year.

Table 7. List of primary studies used in the SLR.

| No. | Title | Year | Venue type | Ref. |
|---|---|---|---|---|
| P1 | Experiment on the automatic detection of function clones in a software system using metrics | 1996 | Conference | [185] |
| P2 | A language independent approach for detecting duplicated code | 1999 | Conference | [90] |
| P3 | Identifying similar code with program dependence graphs | 2001 | Conference | [64] |
| P4 | Using slicing to identify duplication in source code | 2001 | Conference | [186] |
| P5 | CCFinder: a multilinguistic token-based code clone detection system for large scale source code | 2002 | Journal | [2] |
| P6 | Finding plagiarisms among a set of programs with JPlag | 2002 | Journal | [10] |
| P7 | CP-Miner: finding copy-paste and related bugs in large-scale software code | 2006 | Journal | [187] |
| P8 | Detecting similar java classes using tree algorithms | 2006 | Conference | [188] |
| P9 | DECKARD: scalable and accurate tree-based detection of code clones | 2007 | Conference | [82] |
| P10 | Efficient plagiarism detection for large code repositories | 2007 | Journal | [11] |
| P11 | NICAD: accurate detection of near-miss intentional clones using flexible pretty-printing and code normalization | 2008 | Conference | [81] |
| P12 | Scalable detection of semantic clones | 2008 | Conference | [189] |
| P13 | Empirical evaluation of clone detection using syntax suffix trees | 2008 | Journal | [190] |
| P14 | Detection and analysis of near-miss software clones | 2009 | Conference | [191] |
| P15 | Automatic source code plagiarism detection | 2009 | Conference | [98] |
| P16 | A mutation/injection-based automatic framework for evaluating code clone detection tools | 2009 | Conference | [93] |
| P17 | A data mining approach for detecting higher-level clones in software | 2009 | Journal | [165] |
| P18 | CloneDetective - a workbench for clone detection research | 2009 | Conference | [158] |
| P19 | Clone detection via structural abstraction | 2009 | Journal | [192] |

| No. | Title | Year | Venue type | Ref. |
|---|---|---|---|---|
| P20 | Enhancing source-based clone detection using intermediate representation | 2010 | Conference | [193] |
| P21 | Incremental code clone detection: a PDG-based approach | 2011 | Conference | [164] |
| P22 | An approach to source-code plagiarism detection and investigation using latent semantic analysis | 2012 | Journal | [13] |
| P23 | CBCD: cloned buggy code detector | 2012 | Conference | [103] |
| P24 | STVsm: similar structural code detection based on AST and VSM | 2012 | Conference | [44] |
| P25 | An efficient new multi-language clone detection approach from large source code | 2012 | Conference | [194] |
| P26 | A source code similarity system for plagiarism detection | 2013 | Journal | [12] |
| P27 | Semantic code clone detection using parse trees and grammar recovery | 2013 | Conference | [195] |
| P28 | An efficient code clone detection model on Java byte code using hybrid approach | 2013 | Conference | [196] |
| P29 | A hybrid-token and textual based approach to find similar code segments | 2013 | Conference | [197] |
| P30 | Gapped code clone detection with lightweight source code analysis | 2013 | Conference | [198] |
| P31 | CodeBlast: a two-stage algorithm for improved program similarity matching in large software repositories | 2013 | Conference | [159] |
| P32 | CPDP: a robust technique for plagiarism detection in source code | 2013 | Conference | [99] |
| P33 | Structural detection of android malware using embedded call graphs | 2013 | Conference | [100] |
| P34 | SPAPE: A semantic-preserving amorphous procedure extraction method for near-miss clones | 2013 | Journal | [160] |
| P35 | gCad: A near-miss clone genealogy extractor to support clone evolution analysis | 2013 | Conference | [199] |
| P36 | Viewing functions as token sequences to highlight similarities in source code | 2013 | Journal | [151] |
| P37 | Clonewise – detecting package-level clones using machine learning | 2013 | Conference | [200] |
| P38 | How should we measure functional sameness from program source code? An exploratory study on java methods | 2014 | Conference | [201] |
| P39 | Method-level code clone detection through LWH (light weight hybrid) approach | 2014 | Journal | [202] |
| P40 | Detection of semantically similar code | 2014 | Journal | [45] |
| P41 | LLVM-based code clone detection framework | 2015 | Conference | [203] |
| P42 | Threshold-free code clone detection for a large-scale heterogeneous Java repository | 2015 | Conference | [204] |
| P43 | Detecting Android malware using clone detection | 2015 | Journal | [17] |
| P44 | Structural information based malicious app similarity calculation and clustering | 2015 | Conference | [205] |
| P45 | An execution-semantic and content-and-context-based code-clone detection and analysis | 2015 | Conference | [206] |
| P46 | Software clone detection using clustering approach | 2015 | Conference | [207] |
| P47 | SourcererCC: scaling code clone detection to big-code | 2016 | Conference | [153] |
| P48 | Semantic clone detection using machine learning | 2016 | Conference | [177] |
| P49 | Deep learning code fragments for code clone detection | 2016 | Conference | [166] |
| P50 | Identifying functionally similar code in complex codebases | 2016 | Conference | [85] |
| P51 | Measuring source code similarity by finding similar subgraph with an incremental genetic algorithm | 2016 | Conference | [46] |
| P52 | Code relatives: detecting similarly behaving software | 2016 | Conference | [101] |
| P53 | SourcererCC and SourcererCC-I: Tools to detect clones in batch mode and during software development | 2016 | Conference | [154] |
| P54 | Semantic code clone detection for Internet of Things applications using reaching definition and liveness analysis | 2016 | Journal | [41] |
| P55 | Similarity management of 'cloned and owned' variants | 2016 | Conference | [208] |
| P56 | Code clone detection based on order and content of control statements | 2016 | Conference | [209] |

| No. | Title | Year | Venue type | Ref. |
|---|---|---|---|---|
| P57 | Discovering vulnerable functions: a code similarity based approach | 2016 | Conference | [210] |
| P58 | CCSharp: an efficient three-phase code clone detection using modified PDGs | 2017 | Conference | [83] |
| P59 | CCLearner: a deep learning-based clone detection approach | 2017 | Conference | [211] |
| P60 | GroupDroid: automatically grouping mobile malware by extracting code similarities | 2017 | Conference | [18] |
| P61 | DéjàVu: a map of code duplicates on GitHub | 2017 | Conference | [212] |
| P62 | A case study of TTCN-3 test scripts clone analysis in an industrial telecommunication setting | 2017 | Journal | [213] |
| P63 | Using compilation/decompilation to enhance clone detection | 2017 | Conference | [214] |
| P64 | SCVD: a new semantics-based approach for cloned vulnerable code detection | 2017 | Conference | [215] |
| P65 | Deep learning similarities from different representations of source code | 2018 | Conference | [216] |
| P66 | Test-based clone detection: an initial try on semantically equivalent methods | 2018 | Journal | [84] |
| P67 | On the use of machine learning techniques towards the design of cloud based automatic code clone validation tools | 2018 | Conference | [217] |
| P68 | CCAligner: a token based large-gap clone detector | 2018 | Conference | [157] |
| P69 | DroidCC: a scalable clone detection approach for android applications to detect similarity at source code level | 2018 | Conference | [47] |
| P70 | Structural function based code clone detection using a new hybrid technique | 2018 | Conference | [59] |
| P71 | Detecting functionally similar code within the same project | 2018 | Conference | [218] |
| P72 | Oreo: detection of clones in the twilight zone | 2018 | Conference | [40] |
| P73 | DeepSim: deep learning code functional similarity | 2018 | Conference | [147] |
| P74 | Tackling Androids native library malware with robust, efficient and accurate similarity measures | 2018 | Conference | [19] |
| P75 | Plagiarism detection in students' programming assignments based on semantics: multimedia e-learning based smart assessment methodology | 2018 | Journal | [15] |
| P76 | LICCA: A tool for cross-language clone detection | 2018 | Conference | [219] |
| P77 | Image-based clone code detection and visualization | 2019 | Conference | [220] |
| P78 | TECCD: a tree embedding approach for code clone detection | 2019 | Conference | [221] |
| P79 | A large-gap clone detection approach using sequence alignment via dynamic parameter optimization | 2019 | Journal | [112] |
| P80 | Fast code clone detection based on weighted recursive autoencoders | 2019 | Journal | [222] |
| P81 | Cross-language clone detection by learning over abstract syntax trees | 2019 | Conference | [149] |
| P82 | A novel neural source code representation based on abstract syntax tree | 2019 | Conference | [223] |
| P83 | Learning-based recursive aggregation of abstract syntax trees for code clone detection | 2019 | Conference | [7] |
| P84 | CloneCognition: machine learning based code clone validation tool | 2019 | Conference | [224] |
| P85 | Code2vec: learning distributed representations of code | 2019 | Conference | [23] |
| P86 | Code similarity detection through control statement and program features | 2019 | Journal | [225] |
| P87 | Siamese: scalable and incremental code clone search via multiple code representations | 2019 | Journal | [152] |
| P88 | Design and development of software tool for code clone search, detection, and analysis | 2019 | Conference | [226] |
| P89 | Comparison of token-based code clone method with pattern mining technique and traditional string matching algorithms in- terms of software reuse | 2019 | Conference | [49] |
| P90 | A function level Java code clone detection method | 2019 | Conference | [227] |
| P91 | Cross-project code clones in GitHub | 2019 | Journal | [228] |
| P92 | SCDetector: software functional clone detection based on semantic tokens analysis | 2020 | Conference | [156] |
| P93 | SAGA: efficient and large-scale detection of near-miss clones with GPU acceleration | 2020 | Conference | [229] |

| No. | Title | Year | Venue type | Ref. |
|-----|-------|------|------------|------|
| P94 | From local to global semantic clone detection | 2020 | Conference | [230] |
| P95 | CPPCD: a token-based approach to detecting potential clones | 2020 | Conference | [155] |
| P96 | Improving syntactical clone detection methods through the use of an intermediate representation | 2020 | Conference | [231] |
| P97 | Twin-Finder: integrated reasoning engine for pointer-related code clone detection | 2020 | Conference | [232] |
| P98 | LVMapper: a large-variance clone detector using sequencing alignment approach | 2020 | Journal | [233] |
| P99 | Functional code clone detection with syntax and semantics fusion learning | 2020 | Conference | [9] |
| P100 | CLCDSA: cross language code clone detection using syntactical features and API documentation | 2020 | Conference | [61] |
| P101 | A universal cross language software similarity detector for open source software categorization | 2020 | Journal | [60] |
| P102 | A machine learning based framework for code clone validation | 2020 | Journal | [234] |
| P103 | DeepClone: modeling clones to generate code predictions | 2020 | Conference | [167] |
| P104 | A novel code stylometry-based code clone detection strategy | 2020 | Conference | [235] |
| P105 | Learn to align: a code alignment network for code clone detection | 2021 | Conference | [236] |
| P106 | You look so different: finding structural clones and subclones in java source code | 2021 | Conference | [237] |
| P107 | InferCode: self-supervised learning of code representations by predicting subtrees | 2021 | Conference | [238] |
| P108 | CCEyes: an effective tool for code clone detection on large-scale open source repositories | 2021 | Conference | [168] |
| P109 | FCCA: hybrid code representation for functional clone detection using attention networks | 2021 | Journal | [239] |
| P110 | CCGraph: a PDG-based code clone detector with approximate graph matching | 2021 | Conference | [161] |
| P111 | Stubber: Compiling source code into bytecode without dependencies for Java code clone Detection | 2021 | Conference | [240] |
| P112 | An effective semantic code clone detection framework using pairwise feature fusion | 2021 | Journal | [241] |
| P113 | Development of porting analyzer to search cross-language code clones using Levenshtein distance | 2021 | Conference | [242] |
| P114 | Improving code clone detection accuracy and efficiency based on code complexity analysis | 2022 | Conference | [243] |
| P115 | Cross-project software defect prediction based on class code similarity | 2022 | Journal | [102] |
| P116 | C4: contrastive cross-language code clone detection | 2022 | Conference | [244] |
| P117 | Unleashing the power of compiler intermediate representation to enhance neural program embeddings | 2022 | Conference | [245] |
| P118 | Clone-Seeker: effective code clone search using annotations | 2022 | Journal | [246] |
| P119 | Modeling functional similarity in source code with graph-based Siamese networks | 2022 | Journal | [162] |
| P120 | A collaborative method for code clone detection using a deep learning model | 2022 | Journal | [247] |
| P121 | ASTENS-BWA: searching partial syntactic similar regions between source code fragments via AST-based encoded sequence alignment | 2022 | Journal | [104] |
| P122 | Using a nearest-neighbour, BERT-Based approach for scalable clone detection | 2022 | Conference | [248] |
| P123 | Towards overcoming type limitations in semantic clone detection | 2022 | Conference | [249] |
| P124 | Predicting buggy code clones through machine learning | 2022 | Conference | [250] |
| P125 | TreeCen: Building tree graph for scalable semantic code clone detection | 2022 | Conference | [251] |
| P126 | Detecting semantic code clones by building AST-Based Markov chains model | 2022 | Conference | [252] |
| P127 | Semantic clone detection via probabilistic software modeling | 2022 | Conference | [253] |
| P128 | FastDCF: a partial index based distributed and scalable near-miss code clone detection approach for very large code repositories | 2022 | Conference | [254] |

| No. | Title | Year | Venue type | Ref. |
|-----|-------|------|-----------|------|
| P129 | Combining holistic source code representation with Siamese neural networks for detecting code clones | 2022 | Conference | [255] |
| P130 | SEED: Semantic graph based deep detection for type-4 clone | 2022 | Conference | [163] |
| P131 | Precise code clone detection with architecture of abstract syntax trees | 2022 | Conference | [256] |
| P132 | TPCaps: a framework for code clone detection and localization based on improved CapsNet | 2022 | Journal | [257] |
| P133 | Method name recommendation based on source code metrics | 2023 | Journal | [26] |
| P134 | Efficient transformer with code token learner for code clone detection | 2023 | Journal | [258] |
| P135 | Ranking code clones to support maintenance activities | 2023 | Journal | [259] |
| P136 | CCStokener: fast yet accurate code clone detection with semantic token | 2023 | Journal | [260] |

# References

[1]    I. D. Baxter, A. Yahin, L. Moura, M. Sant'Anna, and L. Bier, "Clone detection using abstract syntax trees," in *Proceedings. International Conference on Software Maintenance (Cat. No. 98CB36272)*, IEEE Comput. Soc, 1998, pp. 368–377. doi: 10.1109/ICSM.1998.738528.

[2]    T. Kamiya, S. Kusumoto, and K. Inoue, "CCFinder: a multilinguistic token-based code clone detection system for large scale source code," *IEEE Transactions on Software Engineering*, vol. 28, no. 7, pp. 654–670, Jul. 2002, doi: 10.1109/TSE.2002.1019480.

[3]    S. Carter, R. J. Frank, and D. S. W. Tansley, "Clone detection in telecommunications software systems: a neural net approach," in *Proc. Int. Workshop on Application of Neural Networks to Telecommunications*, 1993, pp. 273–287.

[4]    H. Nasirloo and F. Azimzadeh, "Semantic code clone detection using abstract memory states and program dependency graphs," *2018 4th International Conference on Web Research, ICWR 2018*, pp. 19–27, 2018, doi: 10.1109/ICWR.2018.8387232.

[5]    Roopam and G. Singh, "To enhance the code clone detection algorithm by using hybrid approach for detection of code clones," *Proceedings of the 2017 International Conference on Intelligent Computing and Control Systems, ICICCS 2017*, vol. 2018-Janua, pp. 192–198, 2017, doi: 10.1109/ICCONS.2017.8250708.

[6]    M. R. H. Misu and K. Sakib, "Interface driven code clone detection," *Proceedings - Asia-Pacific Software Engineering Conference, APSEC*, vol. 2017-Decem, pp. 747–748, 2018, doi: 10.1109/APSEC.2017.97.

[7]    L. Buch and A. Andrzejak, "Learning-based recursive aggregation of abstract syntax trees for code clone detection," *SANER 2019 - Proceedings of the 2019 IEEE 26th International Conference on Software Analysis, Evolution, and Reengineering*, pp. 95–104, 2019, doi: 10.1109/SANER.2019.8668039.

[8]    R. Koschke, R. Falke, and P. Frenzel, "Clone detection using abstract syntax suffix trees," in *2006 13th Working Conference on Reverse Engineering*, IEEE, 2006, pp. 253–262. doi: 10.1109/WCRE.2006.18.

[9]    C. Fang, Z. Liu, Y. Shi, J. Huang, and Q. Shi, "Functional code clone detection with syntax and semantics fusion learning," *ISSTA 2020 - Proceedings of the 29th ACM SIGSOFT International Symposium on Software Testing and Analysis*, pp. 516–527, 2020, doi: 10.1145/3395363.3397362.

[10]    L. Prechelt, G. Malpohl, M. Philippsen, and others, "Finding plagiarisms among a set of programs with JPlag.," *J. Univers. Comput. Sci.*, vol. 8, no. 11, p. 1016, 2002.

[11]    S. Burrows, S. M. M. Tahaghoghi, and J. Zobel, "Efficient plagiarism detection for large code repositories," *Softw Pract Exp*, vol. 37, no. 2, pp. 151–175, Feb. 2007, doi: 10.1002/spe.750.

[12]    Z. Duric and D. Gasevic, "A source code similarity system for plagiarism detection," *Comput J*, vol. 56, no. 1, pp. 70–86, Jan. 2013, doi: 10.1093/comjnl/bxs018.

[13]    G. Cosma and M. Joy, "An approach to source-code plagiarism detection and investigation using latent semantic analysis," *IEEE Transactions on Computers*, vol. 61, no. 3, pp. 379–394, Mar. 2012, doi: 10.1109/TC.2011.223.

[14]    T. Foltýnek, N. Meuschke, and B. Gipp, "Academic plagiarism detection: A systematic literature review," *ACM Comput Surv*, vol. 52, no. 6, 2019, doi: 10.1145/3345317.

[15]    F. Ullah, J. Wang, M. Farhan, S. Jabbar, Z. Wu, and S. Khalid, "Plagiarism detection in students' programming assignments based on semantics: multimedia e-learning based smart assessment methodology," *Multimed Tools Appl*, vol. 79, no. 13–14, pp. 8581–8598, 2020, doi: 10.1007/s11042-018-5827-6.

[16]    C. Liu, C. Chen, J. Han, and P. S. Yu, "GPLAG: detection of software plagiarism by program dependence graph analysis," in *Proceedings of the 12th ACM SIGKDD international conference on Knowledge discovery and data mining - KDD '06*, New York, New York, USA: ACM Press, 2006, p. 872. doi: 10.1145/1150402.1150522.

[17]    J. Chen, M. H. Alalfi, T. R. Dean, and Y. Zou, "Detecting Android malware using clone detection," *J Comput Sci Technol*, vol. 30, no. 5, pp. 942–956, 2015, doi: 10.1007/s11390-015-1573-7.

[18]    N. Marastoni, A. Continella, D. Quarta, S. Zanero, and M. D. Preda, "Groupdroid: Automatically grouping mobile malware by extracting code similarities," *ACM International Conference Proceeding Series*, 2017, doi: 10.1145/3151137.3151138.

[19]    A. Kalysch, M. Protsenko, O. Milisterfer, and T. Müller, "Tackling androids native library malware with robust, efficient and accurate similarity measures," *ACM International Conference Proceeding Series*, 2018, doi: 10.1145/3230833.3232828.

[20]    J. Kim and B.-R. Moon, "New malware detection system using metric-based method and hybrid genetic algorithm," in *Proceedings of the fourteenth international conference on Genetic and evolutionary computation conference companion - GECCO Companion '12*, New York, New York, USA: ACM Press, 2012, p. 1527. doi: 10.1145/2330784.2331029.

[21]    A. M. Lajevardi, S. Parsa, and M. J. Amiri, "Markhor: malware detection using fuzzy similarity of system call dependency sequences," *Journal of Computer Virology and Hacking Techniques*, Apr. 2021, doi: 10.1007/s11416-021-00383-1.

[22]    F. P. Viertel, W. Brunotte, D. Strüber, and K. Schneider, "Detecting security vulnerabilities using clone detection and community knowledge," *Proceedings of the International Conference on Software Engineering and Knowledge Engineering, SEKE*, vol. 2019-July, no. August, pp. 245–252, 2019, doi: 10.18293/SEKE2019-183.

[23]    U. Alon, M. Zilberstein, O. Levy, and E. Yahav, "Code2vec: learning distributed representations of code," in *Proceedings of the ACM on Programming Languages*, Jan. 2019, pp. 1–29. doi: 10.1145/3290353.

[24]    U. Alon, S. Brody, O. Levy, and E. Yahav, "Code2seq: generating sequences from structured representations of code," in *International Conference on Learning Representations*, 2019. Accessed: Sep. 26, 2022. [Online]. Available: https://openreview.net/forum?id=H1gKYo09tX

[25]    S. Arshad, S. Abid, and S. Shamail, "CodeBERT for code clone detection: a replication study," in *2022 IEEE 16th International Workshop on Software Clones (IWSC)*, IEEE, Oct. 2022, pp. 39–45. doi: 10.1109/IWSC55060.2022.00015.

[26]    S. Parsa, M. Zakeri-Nasrabadi, M. Ekhtiarzadeh, and M. Ramezani, "Method name recommendation based on source code metrics," *J Comput Lang*, vol. 74, p. 101177, Jan. 2023, doi: 10.1016/j.cola.2022.101177.

[27]    M. Zakeri-Nasrabadi and S. Parsa, "Learning to predict test effectiveness," *International Journal of Intelligent Systems*, Oct. 2021, doi: 10.1002/int.22722.

[28]    M. Z. Nasrabadi and S. Parsa, "Learning to predict software testability," in *2021 26th International Computer Conference, Computer Society of Iran (CSICC)*, Tehran: IEEE, Mar. 2021, pp. 1–5. doi: 10.1109/CSICC52343.2021.9420548.

[29]    M. Zakeri-Nasrabadi and S. Parsa, "An ensemble meta-estimator to predict source code testability," *Appl Soft Comput*, vol. 129, p. 109562, Nov. 2022, doi: 10.1016/j.asoc.2022.109562.

[30]    M. D. Papamichail, T. Diamantopoulos, and A. L. Symeonidis, "Measuring the reusability of software components using static analysis metrics and reuse rate information," *Journal of Systems and Software*, vol. 158, p. 110423, Dec. 2019, doi: 10.1016/j.jss.2019.110423.

[31]    F. Arcelli Fontana and M. Zanoni, "Code smell severity classification using machine learning techniques," *Knowl Based Syst*, vol. 128, pp. 43–58, Jul. 2017, doi: 10.1016/j.knosys.2017.04.014.

[32]    F. A. Fontana, M. Zanoni, A. Marino, and M. v. Mäntylä, "Code smell detection: towards a machine learning-based approach," *IEEE International Conference on Software Maintenance, ICSM*, pp. 396–399, 2013, doi: 10.1109/ICSM.2013.56.

[33]    F. Arcelli Fontana, M. V. Mäntylä, M. Zanoni, and A. Marino, "Comparing and experimenting machine learning techniques for code smell detection," *Empir Softw Eng*, vol. 21, no. 3, pp. 1143–1191, Jun. 2016, doi: 10.1007/s10664-015-9378-4.

[34]    H. Liu, J. Jin, Z. Xu, Y. Bu, Y. Zou, and L. Zhang, "Deep learning based code smell detection," *IEEE Transactions on Software Engineering*, pp. 1–1, 2021, doi: 10.1109/TSE.2019.2936376.

[35]    M. Aniche, E. Maziero, R. Durelli, and V. Durelli, "The effectiveness of supervised machine learning algorithms in predicting software refactoring," *IEEE Transactions on Software Engineering*, pp. 1–1, 2020, doi: 10.1109/TSE.2020.3021736.

[36]    A. M. Sheneamer, "An automatic advisor for refactoring software clones based on machine learning," *IEEE Access*, vol. 8, pp. 124978–124988, 2020, doi: 10.1109/ACCESS.2020.3006178.

[37]    M. White, M. Tufano, M. Martinez, M. Monperrus, and D. Poshyvanyk, "Sorting and transforming program repair ingredients via deep learning code similarities," in *2019 IEEE 26th International Conference on Software Analysis, Evolution and Reengineering (SANER)*, IEEE, Feb. 2019, pp. 479–490. doi: 10.1109/SANER.2019.8668043.

[38]    H. Cao, F. Liu, J. Shi, Y. Chu, and M. Deng, "Random search and code similarity-based automatic program repair," *J Shanghai Jiaotong Univ Sci*, Nov. 2022, doi: 10.1007/s12204-022-2514-6.

[39]    M. Allamanis, H. Peng, and C. Sutton, "A convolutional attention network for extreme summarization of source code," in *Proceedings of The 33rd International Conference on Machine Learning*, M. F. Balcan and K. Q. Weinberger, Eds., in Proceedings of Machine Learning Research, vol. 48. New York, New York, USA: PMLR, Dec. 2016, pp. 2091–2100. [Online]. Available: https://proceedings.mlr.press/v48/allamanis16.html

[40]    V. Saini, F. Farmahinifarahani, Y. Lu, P. Baldi, and C. V. Lopes, "Oreo: detection of clones in the twilight zone," *ESEC/FSE 2018 - Proceedings of the 2018 26th ACM Joint Meeting on European Software Engineering Conference and Symposium on the Foundations of Software Engineering*, pp. 354–365, 2018, doi: 10.1145/3236024.3236026.

[41]    R. Tekchandani, R. Bhatia, and M. Singh, "Semantic code clone detection for Internet of Things applications using reaching definition and liveness analysis," *Journal of Supercomputing*, vol. 74, no. 9, pp. 4199–4226, 2018, doi: 10.1007/s11227-016-1832-6.

[42]    I. Reinhartz-Berger and A. Zamansky, "Reuse of similarly behaving software through polymorphism-inspired variability mechanisms," *IEEE Transactions on Software Engineering*, vol. 48, no. 3, pp. 773–785, Mar. 2022, doi: 10.1109/TSE.2020.3001512.

[43]    C. Ragkhitwetsagul, J. Krinke, and D. Clark, "A comparison of code similarity analysers," *Empir Softw Eng*, vol. 23, no. 4, pp. 2464–2519, Aug. 2018, doi: 10.1007/s10664-017-9564-7.

[44]    N. Li, M. Shen, S. Li, L. Zhang, and Z. Li, "STVsm: Similar Structural Code Detection Based on AST and VSM," 2012, pp. 15–21. doi: 10.1007/978-3-642-35267-6_3.

[45]    T. Wang, K. Wang, X. Su, and P. Ma, "Detection of semantically similar code," *Front Comput Sci*, vol. 8, no. 6, pp. 996–1011, Dec. 2014, doi: 10.1007/s11704-014-3430-1.

[46]    J. Kim, H. Choi, H. Yun, and B.-R. Moon, "Measuring Source Code Similarity by Finding Similar Subgraph with an Incremental Genetic Algorithm," in *Proceedings of the Genetic and Evolutionary Computation Conference 2016*, New York, NY, USA: ACM, Jul. 2016, pp. 925–932. doi: 10.1145/2908812.2908870.

[47]    J. Akram, Z. Shi, M. Mumtaz, and P. Luo, "DroidCC: a scalable clone detection approach for android applications to detect similarity at source code level," in *2018 IEEE 42nd Annual Computer Software and Applications Conference (COMPSAC)*, IEEE, Jul. 2018, pp. 100–105. doi: 10.1109/COMPSAC.2018.00021.

[48]    C. K. Roy and J. R. Cordy, "Survey on software clone detection research," 2007.

[49]    K. E. Rajakumari, "Comparison of token-based code clone method with pattern mining technique and traditional string matching algorithms in-terms of software reuse," *Proceedings of 2019 3rd IEEE International Conference on Electrical, Computer and Communication Technologies, ICECCT 2019*, pp. 1–6, 2019, doi: 10.1109/ICECCT.2019.8869324.

[50] M. S. Rahman and C. K. Roy, "A change-type based empirical study on the stability of cloned code," *Proceedings - 2014 14th IEEE International Working Conference on Source Code Analysis and Manipulation, SCAM 2014*, pp. 31–40, 2014, doi: 10.1109/SCAM.2014.13.

[51] Z. Á. Mann, "Three public enemies: cut, copy, and paste," *Computer (Long Beach Calif)*, vol. 39, no. 7, pp. 31–35, Jul. 2006, doi: 10.1109/MC.2006.246.

[52] J. H. Johnson, "Substring matching for clone detection and change tracking," in *Proceedings - 1994 International Conference on Software Maintenance, ICSM 1994*, Institute of Electrical and Electronics Engineers Inc., 1994, pp. 120–126. doi: 10.1109/ICSM.1994.336783.

[53] N. Tsantalis, D. Mazinanian, and G. P. Krishnan, "Assessing the refactorability of software clones," *IEEE Transactions on Software Engineering*, vol. 41, no. 11, pp. 1055–1090, Nov. 2015, doi: 10.1109/TSE.2015.2448531.

[54] Y. Higo, T. Kamiya, S. Kusumoto, and K. Inoue, "Refactoring support based on code clone analysis," 2004, pp. 220–233. doi: 10.1007/978-3-540-24659-6_16.

[55] M. Fowler and K. Beck, *Refactoring: improving the design of existing code*, Second Edi. Addison-Wesley, 2018. [Online]. Available: https://refactoring.com/

[56] T. Yamamoto, M. Matsushita, T. Kamiya, and K. Inoue, "Measuring similarity of large software systems based on source code correspondence," 2005, pp. 530–544. doi: 10.1007/11497455_41.

[57] L. Qinqin and Z. Chunhai, "Research on algorithm of program code similarity detection," in *2017 International Conference on Computer Systems, Electronics and Control (ICCSEC)*, 2017, pp. 1289–1292.

[58] S. Giesecke, "Generic modelling of code clones," in *Duplication, Redundancy, and Similarity in Software*, 2006.

[59] Y. Yang, Z. Ren, X. Chen, and H. Jiang, "Structural function based code clone detection using a new hybrid technique," *Proceedings - International Computer Software and Applications Conference*, vol. 1, pp. 286–291, 2018, doi: 10.1109/COMPSAC.2018.00045.

[60] K. W. Nafi, B. Roy, C. K. Roy, and K. A. Schneider, "A universal cross language software similarity detector for open source software categorization," *Journal of Systems and Software*, vol. 162, p. 110491, 2020, doi: 10.1016/j.jss.2019.110491.

[61] K. W. Nafi, T. S. Kar, B. Roy, C. K. Roy, and K. A. Schneider, "CLCDSA: cross language code clone detection using syntactical features and API documentation," *Proceedings - 2019 34th IEEE/ACM International Conference on Automated Software Engineering, ASE 2019*, pp. 1026–1037, 2019, doi: 10.1109/ASE.2019.00099.

[62] S. Bellon, R. Koschke, G. Antoniol, J. Krinke, and E. Merlo, "Comparison and evaluation of clone detection tools," *IEEE Transactions on Software Engineering*, vol. 33, no. 9, pp. 577–591, Sep. 2007, doi: 10.1109/TSE.2007.70725.

[63] N. Davey, P. Barson, S. Field, R. Frank, and D. Tansley, "The development of a software clone detector," *International Journal of Applied Software Technology*, 1995.

[64] J. Krinke, "Identifying similar code with program dependence graphs," in *Proceedings Eighth Working Conference on Reverse Engineering*, IEEE Comput. Soc, 2001, pp. 301–309. doi: 10.1109/WCRE.2001.957835.

[65] G. Mishne and M. de Rijke, "Source code retrieval using conceptual similarity," in *Coupling Approaches, Coupling Media and Coupling Languages for Information Retrieval*, in RIAO '04. Paris, FRA: LE CENTRE DE HAUTES ETUDES INTERNATIONALES D'INFORMATIQUE DOCUMENTAIRE, 2004, pp. 539–554.

[66] A. Lakhotia, Junwei Li, A. Walenstein, and Yun Yang, "Towards a clone detection benchmark suite and results archive," in *MHS2003. Proceedings of 2003 International Symposium on Micromechatronics and Human Science (IEEE Cat. No.03TH8717)*, IEEE Comput. Soc, pp. 285–286. doi: 10.1109/WPC.2003.1199215.

[67] D. Rattan, R. Bhatia, and M. Singh, *Software clone detection: a systematic review*, vol. 55, no. 7. Elsevier B.V., 2013. doi: 10.1016/j.infsof.2013.01.008.

[68] H. Min and Z. Li Ping, "Survey on software clone detection research," in *Proceedings of the 2019 3rd International Conference on Management Engineering, Software Engineering and Service Sciences - ICMSS 2019*, New York, New York, USA: ACM Press, 2019, pp. 9–16. doi: 10.1145/3312662.3312707.

[69] Q. U. Ain, W. H. Butt, M. W. Anwar, F. Azam, and B. Maqbool, "A systematic review on code clone detection," *IEEE Access*, vol. 7, pp. 86121–86144, 2019, doi: 10.1109/ACCESS.2019.2918202.

[70] C.-F. Chen, A. M. Zain, and K.-Q. Zhou, "Definition, approaches, and analysis of code duplication detection (2006–2020): a critical review," *Neural Comput Appl*, vol. 34, no. 23, pp. 20507–20537, Dec. 2022, doi: 10.1007/s00521-022-07707-2.

[71] M. Lei, H. Li, J. Li, N. Aundhkar, and D.-K. Kim, "Deep learning application on code clone detection: a review of current knowledge," *Journal of Systems and Software*, vol. 184, p. 111141, Feb. 2022, doi: 10.1016/j.jss.2021.111141.

[72] M. Novak, M. Joy, and D. Kermek, "Source-code similarity detection and detection tools used in academia: A systematic review," *ACM Transactions on Computing Education*, vol. 19, no. 3, 2019, doi: 10.1145/3313290.

[73] E. Burd and J. Bailey, "Evaluating clone detection tools for use during preventative maintenance," in *Proceedings. Second IEEE International Workshop on Source Code Analysis and Manipulation*, IEEE Comput. Soc, pp. 36–43. doi: 10.1109/SCAM.2002.1134103.

[74] B. Biegel, Q. D. Soetens, W. Hornig, S. Diehl, and S. Demeyer, "Comparison of similarity metrics for refactoring detection," in *Proceeding of the 8th working conference on Mining software repositories - MSR '11*, New York, New York, USA: ACM Press, 2011, p. 53. doi: 10.1145/1985441.1985452.

[75] B. Kitchenham and S. Charters, "Guidelines for performing systematic literature reviews in software engineering," 2007.

[76] A. S. Nuñez-Varela, H. G. Pérez-Gonzalez, F. E. Martínez-Perez, and C. Soubervielle-Montalvo, "Source code metrics: a systematic mapping study," *Journal of Systems and Software*, vol. 128, pp. 164–197, 2017, doi: https://doi.org/10.1016/j.jss.2017.03.044.

[77] M. I. Azeem, F. Palomba, L. Shi, and Q. Wang, "Machine learning techniques for code smell detection: a systematic literature review and meta-analysis," *Inf Softw Technol*, vol. 108, pp. 115–138, Apr. 2019, doi: 10.1016/j.infsof.2018.12.009.

[78] C. Abid, V. Alizadeh, M. Kessentini, T. do N. Ferreira, and D. Dig, "30 years of software refactoring research: a systematic literature review," *arXiv preprint arXiv:2007.02194*, 2020.

[79] C. Wohlin, "Guidelines for snowballing in systematic literature studies and a replication in software engineering," in *Proceedings of the 18th International Conference on Evaluation and Assessment in Software Engineering - EASE '14*, New York, New York, USA: ACM Press, 2014, pp. 1–10. doi: 10.1145/2601248.2601268.

[80] M. Zakeri-Nasrabadi, S. Parsa, M. Ramezani, C. Roy, and M. Ekhtiarzadeh, "Supplementary data for a systematic literature review on source code similarity measurement and clone detection: techniques, applications, and challenges," May 31, 2023. https://doi.org/10.5281/zenodo.7993619 (accessed May 31, 2023).

[81] C. K. Roy and J. R. Cordy, "NICAD: accurate detection of near-miss intentional clones using flexible pretty-printing and code normalization," in *2008 16th IEEE International Conference on Program Comprehension*, IEEE, Jun. 2008, pp. 172–181. doi: 10.1109/ICPC.2008.41.

[82] L. Jiang, G. Misherghi, Z. Su, and S. Glondu, "Deckard: scalable and accurate tree-based detection of code clones," in *29th International Conference on Software Engineering (ICSE'07)*, 2007, pp. 96–105.

[83] M. Wang, P. Wang, and Y. Xu, "CCSharp: an efficient three-phase code clone detector using modified PDGs," *Proceedings - Asia-Pacific Software Engineering Conference, APSEC*, vol. 2017-Decem, pp. 100–109, 2018, doi: 10.1109/APSEC.2017.16.

[84] G. Li, H. Liu, Y. Jiang, and J. Jin, "Test-Based clone detection: an initial try on semantically equivalent methods," *IEEE Access*, vol. 6, pp. 77643–77655, 2018, doi: 10.1109/ACCESS.2018.2883699.

[85] Fang-Hsiang Su, J. Bell, G. Kaiser, and S. Sethumadhavan, "Identifying functionally similar code in complex codebases," in *2016 IEEE 24th International Conference on Program Comprehension (ICPC)*, IEEE, May 2016, pp. 1–10. doi: 10.1109/ICPC.2016.7503720.

[86] "PMD: an extensible cross-language static code analyzer." https://pmd.github.io/ (accessed Sep. 21, 2021).

[87] softwareclones.org, "iClones: incremental clone detection." http://www.softwareclones.org/iclones.php (accessed Dec. 22, 2022).

[88] Daniel Lochner, Joshua Lochner, and Carl Combrinck, "AutoMOSS." https://github.com/automoss/automoss (accessed Dec. 15, 2022).

[89]    Quandary Peak Research, "Simian - similarity analyser." https://simian.quandarypeak.com/ (accessed Dec. 15, 2022).

[90]    S. Ducasse, M. Rieger, and S. Demeyer, "A language independent approach for detecting duplicated code," in *Proceedings IEEE International Conference on Software Maintenance - 1999 (ICSM'99). "Software Maintenance for Business Change" (Cat. No.99CB36360)*, IEEE, 1999, pp. 109–118. doi: 10.1109/ICSM.1999.792593.

[91]    A. Majd, M. Vahidi-Asl, A. Khalilian, A. Baraani-Dastjerdi, and B. Zamani, "Code4Bench: a multidimensional benchmark of Codeforces data for different program analysis techniques," *J Comput Lang*, vol. 53, pp. 38–52, Aug. 2019, doi: 10.1016/j.cola.2019.03.006.

[92]    J. Svajlenko and C. K. Roy, "BigCloneEval: A clone detection tool evaluation framework with BigCloneBench," in *2016 IEEE International Conference on Software Maintenance and Evolution (ICSME)*, IEEE, Oct. 2016, pp. 596–600. doi: 10.1109/ICSME.2016.62.

[93]    C. K. Roy and J. R. Cordy, "A mutation/Injection-based automatic framework for evaluating code clone detection tools," in *2009 International Conference on Software Testing, Verification, and Validation Workshops*, IEEE, 2009, pp. 157–166. doi: 10.1109/ICSTW.2009.18.

[94]    Google, "Google Code Jam," 2016. https://code.google.com/codejam/contests.html (accessed Nov. 17, 2021).

[95]    Mike Mirzayanov, "Codeforces: the only programming contests web 2.0 platform," 2022. https://codeforces.com/ (accessed Dec. 11, 2022).

[96]    S. Yu, T. Wang, and J. Wang, "Data augmentation by program transformation," *Journal of Systems and Software*, vol. 190, p. 111304, Aug. 2022, doi: 10.1016/j.jss.2022.111304.

[97]    T. Lavoie, M. Mérineau, E. Merlo, and P. Potvin, "A case study of TTCN-3 test scripts clone analysis in an industrial telecommunication setting," *Inf Softw Technol*, vol. 87, pp. 32–45, 2017, doi: 10.1016/j.infsof.2017.01.008.

[98]    C. Kustanto and I. Liem, "Automatic source code plagiarism detection," in *2009 10th ACIS International Conference on Software Engineering, Artificial Intelligences, Networking and Parallel/Distributed Computing*, IEEE, May 2009, pp. 481–486. doi: 10.1109/SNPD.2009.62.

[99]    B. Muddu, A. Asadullah, and V. Bhat, "CPDP: a robust technique for plagiarism detection in source code," in *2013 7th International Workshop on Software Clones (IWSC)*, IEEE, May 2013, pp. 39–45. doi: 10.1109/IWSC.2013.6613041.

[100]   H. Gascon, F. Yamaguchi, D. Arp, and K. Rieck, "Structural detection of Android malware using embedded call graphs," *Proceedings of the ACM Conference on Computer and Communications Security*, pp. 45–54, 2013, doi: 10.1145/2517312.2517315.

[101]   F.-H. Su, J. Bell, K. Harvey, S. Sethumadhavan, G. Kaiser, and T. Jebara, "Code relatives: detecting similarly behaving software," in *Proceedings of the 2016 24th ACM SIGSOFT International Symposium on Foundations of Software Engineering*, New York, NY, USA: ACM, Nov. 2016, pp. 702–714. doi: 10.1145/2950290.2950321.

[102]   W. Wen *et al.*, "Cross-project software defect prediction based on class code similarity," *IEEE Access*, vol. 10, pp. 105485–105495, 2022, doi: 10.1109/ACCESS.2022.3211401.

[103]   J. Li and M. D. Ernst, "CBCD: cloned buggy code detector," in *2012 34th International Conference on Software Engineering (ICSE)*, IEEE, Jun. 2012, pp. 310–320. doi: 10.1109/ICSE.2012.6227183.

[104]   Y. Yu, Z. Huang, G. Shen, W. Li, and Y. Shao, "ASTENS-BWA: searching partial syntactic similar regions between source code fragments via AST-based encoded sequence alignment," *Sci Comput Program*, vol. 222, p. 102839, Oct. 2022, doi: 10.1016/j.scico.2022.102839.

[105]   H. Yonai, Y. Hayase, and H. Kitagawa, "Mercem: method name recommendation based on call graph embedding," *Proceedings - Asia-Pacific Software Engineering Conference, APSEC*, vol. 2019-Decem, pp. 134–141, 2019, doi: 10.1109/APSEC48747.2019.00027.

[106]   S. Kurimoto, Y. Hayase, H. Yonai, H. Ito, and H. Kitagawa, "Class name recommendation based on graph embedding of program elements," *Proceedings - Asia-Pacific Software Engineering Conference, APSEC*, vol. 2019-Decem, pp. 498–505, 2019, doi: 10.1109/APSEC48747.2019.00073.

[107]   T. F. Smith and M. S. Waterman, "Identification of common molecular subsequences," *J Mol Biol*, vol. 147, no. 1, pp. 195–197, Mar. 1981, doi: 10.1016/0022-2836(81)90087-5.

[108] R. Agrawal and R. Srikant, "Fast algorithms for mining association rules in large databases," in *Proceedings of the 20th International Conference on Very Large Data Bases*, in VLDB '94. San Francisco, CA, USA: Morgan Kaufmann Publishers Inc., 1994, pp. 487–499.

[109] J. R. Cordy and C. K. Roy, "The NiCad clone detector," in *2011 IEEE 19th International Conference on Program Comprehension*, IEEE, Jun. 2011, pp. 219–220. doi: 10.1109/ICPC.2011.26.

[110] D. Jurafsky and J. H. Martin, *Speech and language processing (second edition)*. Upper Saddle River, NJ, USA: Prentice-Hall, Inc., 2009.

[111] M. J. Wise, "String similarity via greedy string tiling and running Karp-Rabin matching," *Online Preprint, Dec*, vol. 119, no. 1, pp. 1–17, 1993.

[112] J. Liu, T. Wang, C. Feng, H. Wang, and D. Li, "A large-gap clone detection approach using sequence alignment via dynamic parameter optimization," *IEEE Access*, vol. 7, pp. 131270–131281, 2019, doi: 10.1109/ACCESS.2019.2940710.

[113] S. T. Dumais, "Latent semantic analysis," *Annual Review of Information Science and Technology*, vol. 38, no. 1, pp. 188–230, Sep. 2005, doi: 10.1002/aris.1440380105.

[114] X. Yan, J. Han, and R. Afshar, "CloSpan: Mining: Closed sequential patterns in large datasets," in *Proceedings of the 2003 SIAM international conference on data mining*, 2003, pp. 166–177.

[115] P. S. Honnutagi, "The Hadoop distributed file system," *International Journal of Computer Science and Information Technologies (IJCSIT)*, vol. 5, no. 5, pp. 6238–6243, 2014.

[116] J. Dean and S. Ghemawat, "MapReduce: a flexible data processing tool," *Commun ACM*, vol. 53, no. 1, pp. 72–77, Jan. 2010, doi: 10.1145/1629175.1629198.

[117] T. Mikolov, I. Sutskever, K. Chen, G. Corrado, and J. Dean, "Distributed representations of words and phrases and their compositionality," pp. 1–9, 2013, doi: 10.1162/jmlr.2003.3.4-5.951.

[118] T. Parr and K. Fisher, "LL(*): the foundation of the ANTLR parser generator," *Proceedings of the 32nd ACM SIGPLAN conference on Programming language design and implementation*, pp. 425–436, 2011, doi: http://doi.acm.org/10.1145/1993498.1993548.

[119] S. Horwitz and T. Reps, "The use of program dependence graphs in software engineering," in *Proceedings of the 14th international conference on Software engineering - ICSE '92*, New York, New York, USA: ACM Press, 1992, pp. 392–411. doi: 10.1145/143062.143156.

[120] C. Lattner and V. Adve, "LLVM: a compilation framework for lifelong program analysis and transformation," in *International Symposium on Code Generation and Optimization, 2004. CGO 2004.*, IEEE, pp. 75–86. doi: 10.1109/CGO.2004.1281665.

[121] G. Salton, A. Wong, and C. S. Yang, "A vector space model for automatic indexing," *Commun ACM*, vol. 18, no. 11, pp. 613–620, Nov. 1975, doi: 10.1145/361219.361220.

[122] M. Harman, "The role of Artificial Intelligence in Software Engineering," in *2012 First International Workshop on Realizing AI Synergies in Software Engineering (RAISE)*, IEEE, Jun. 2012, pp. 1–6. doi: 10.1109/RAISE.2012.6227961.

[123] E. Schubert, J. Sander, M. Ester, H. P. Kriegel, and X. Xu, "DBSCAN Revisited, Revisited: Why and How You Should (Still) Use DBSCAN," *ACM Transactions on Database Systems*, vol. 42, no. 3, pp. 1–21, Sep. 2017, doi: 10.1145/3068335.

[124] D. Chicco, "Siamese neural networks: an overview," 2021, pp. 73–94. doi: 10.1007/978-1-0716-0826-5_3.

[125] M. Kwabena Patrick, A. Felix Adekoya, A. Abra Mighty, and B. Y. Edward, "Capsule Networks – A survey," *Journal of King Saud University - Computer and Information Sciences*, vol. 34, no. 1, pp. 1295–1310, Jan. 2022, doi: 10.1016/J.JKSUCI.2019.09.014.

[126] L. Jiang, H. Liu, and H. Jiang, "Machine learning based recommendation of method names: how far are we," in *2019 34th IEEE/ACM International Conference on Automated Software Engineering (ASE)*, IEEE, Nov. 2019, pp. 602–614. doi: 10.1109/ASE.2019.00062.

[127] O. Zaitsev, S. Ducasse, A. Bergel, and M. Eveillard, "Suggesting descriptive method names: an exploratory study of two machine learning approaches," 2020, pp. 93–106. doi: 10.1007/978-3-030-58793-2_8.

[128] Y. Li, C. Gu, T. Dullien, O. Vinyals, and P. Kohli, "Graph matching networks for learning the similarity of graph structured objects," in *International conference on machine learning*, 2019, pp. 3835–3845.

[129] D. Guo *et al.*, "Graphcodebert: pre-training code representations with data flow," *arXiv preprint arXiv:2009.08366*, 2020.

[130] J. Devlin, M.-W. Chang, K. Lee, and K. Toutanova, "BERT: Pre-training of deep bidirectional transformers for language understanding," in *Proceedings of the 2019 Conference of the North American Chapter of the Association for Computational Linguistics: Human Language Technologies, NAACL-HLT 2019, Minneapolis, MN, USA, June 2-7, 2019, Volume 1 (Long and Short Papers)*, J. Burstein, C. Doran, and T. Solorio, Eds., Association for Computational Linguistics, 2019, pp. 4171–4186. doi: 10.18653/v1/n19-1423.

[131] B. Rozière, M.-A. Lachaux, M. Szafraniec, and G. Lample, "DOBF: a deobfuscation pre-training objective for programming languages," *CoRR*, vol. abs/2102.07492, 2021, [Online]. Available: https://arxiv.org/abs/2102.07492

[132] C. Ragkhitwetsagul, J. Krinke, and B. Marnette, "A picture is worth a thousand words: Code clone detection based on image similarity," *2018 IEEE 12th International Workshop on Software Clones, IWSC 2018 - Proceedings*, vol. 2018-Janua, pp. 44–50, 2018, doi: 10.1109/IWSC.2018.8327318.

[133] G. Fraser and A. Arcuri, "EvoSuite: automatic test suite generation for object-oriented software," in *Proceedings of the 19th ACM SIGSOFT symposium and the 13th European conference on Foundations of software engineering - SIGSOFT/FSE '11*, New York, New York, USA: ACM Press, 2011, p. 416. doi: 10.1145/2025113.2025179.

[134] I. Goodfellow, Y. Bengio, and A. Courville, *Deep learning*. MIT Press, 2016. [Online]. Available: http://www.deeplearningbook.org/

[135] J. Svajlenko, J. F. Islam, I. Keivanloo, C. K. Roy, and M. M. Mia, "Towards a big data curated benchmark of inter-project code clones," in *2014 IEEE International Conference on Software Maintenance and Evolution*, 2014, pp. 476–480.

[136] Alex Aiken, "MOSS: a system for detecting software similarity." https://theory.stanford.edu/~aiken/moss/ (accessed Dec. 15, 2022).

[137] H. Murakami, Y. Higo, and S. Kusumoto, "A dataset of clone references with gaps," in *Proceedings of the 11th Working Conference on Mining Software Repositories - MSR 2014*, New York, New York, USA: ACM Press, 2014, pp. 412–415. doi: 10.1145/2597073.2597133.

[138] A. Charpentier, J.-R. Falleri, D. Lo, and L. Réveillère, "An empirical assessment of Bellon's clone benchmark," in *Proceedings of the 19th International Conference on Evaluation and Assessment in Software Engineering*, New York, NY, USA: ACM, Apr. 2015, pp. 1–10. doi: 10.1145/2745802.2745821.

[139] J. Svajlenko and C. K. Roy, "Evaluating clone detection tools with BigCloneBench," in *2015 IEEE International Conference on Software Maintenance and Evolution (ICSME)*, IEEE, Sep. 2015, pp. 131–140. doi: 10.1109/ICSM.2015.7332459.

[140] J. Krinke and C. Ragkhitwetsagul, "BigCloneBench considered harmful for machine learning," in *2022 IEEE 16th International Workshop on Software Clones (IWSC)*, 2022, pp. 1–7.

[141] H. Wei and M. Li, "Supervised deep features for software functional clone detection by exploiting lexical and syntactical information in source code," in *Proceedings of the Twenty-Sixth International Joint Conference on Artificial Intelligence*, California: International Joint Conferences on Artificial Intelligence Organization, Aug. 2017, pp. 3034–3040. doi: 10.24963/ijcai.2017/423.

[142] A. Schafer, W. Amme, and T. S. Heinze, "Experiments on code clone detection and machine learning," in *2022 IEEE 16th International Workshop on Software Clones (IWSC)*, IEEE, Oct. 2022, pp. 46–52. doi: 10.1109/IWSC55060.2022.00016.

[143] S. Lu *et al.*, "CodeXGLUE: a machine learning benchmark dataset for code understanding and generation," *CoRR*, vol. abs/2102.04664, 2021.

[144] L. Mou, G. Li, L. Zhang, T. Wang, and Z. Jin, "Convolutional neural networks over tree structures for programming language processing," in *Proceedings of the Thirtieth AAAI Conference on Artificial Intelligence*, in AAAI'16. AAAI Press, 2016, pp. 1287–1293.

[145] Z. Xue, Z. Jiang, C. Huang, R. Xu, X. Huang, and L. Hu, "SEED: semantic graph based deep detection for type-4 clone," 2022, pp. 120–137. doi: 10.1007/978-3-031-08129-3_8.

[146] E. Flores, P. Rosso, L. Moreno, and E. Villatoro-Tello, "On the detection of source code re-use," in *Proceedings of the Forum for Information Retrieval Evaluation on - FIRE '14*, New York, New York, USA: ACM Press, 2015, pp. 21–30. doi: 10.1145/2824864.2824878.

[147] G. Zhao and J. Huang, "DeepSim: deep learning code functional similarity," *ESEC/FSE 2018 - Proceedings of the 2018 26th ACM Joint Meeting on European Software Engineering Conference and Symposium on the Foundations of Software Engineering*, pp. 141–151, 2018, doi: 10.1145/3236024.3236068.

[148] "AtCoder." https://atcoder.jp/ (accessed Jun. 01, 2023).

[149] D. Perez and S. Chiba, "Cross-language clone detection by learning over abstract syntax trees," in *2019 IEEE/ACM 16th International Conference on Mining Software Repositories (MSR)*, IEEE, May 2019, pp. 518–528. doi: 10.1109/MSR.2019.00078.

[150] J. Kim, H. G. Choi, H. Yun, and B. R. Moon, "Measuring source code similarity by finding similar subgraph with an incremental genetic algorithm," *GECCO 2016 - Proceedings of the 2016 Genetic and Evolutionary Computation Conference*, pp. 925–932, 2016, doi: 10.1145/2908812.2908870.

[151] M. Chilowicz, É. Duris, and G. Roussel, "Viewing functions as token sequences to highlight similarities in source code," *Sci Comput Program*, vol. 78, no. 10, pp. 1871–1891, 2013, doi: 10.1016/j.scico.2012.11.008.

[152] C. Ragkhitwetsagul and J. Krinke, "Siamese: scalable and incremental code clone search via multiple code representations," *Empir Softw Eng*, vol. 24, no. 4, pp. 2236–2284, 2019, doi: 10.1007/s10664-019-09697-7.

[153] H. Sajnani, V. Saini, J. Svajlenko, C. K. Roy, and C. V. Lopes, "SourcererCC: scaling code clone detection to big-code," in *Proceedings of the 38th International Conference on Software Engineering*, New York, NY, USA: ACM, May 2016, pp. 1157–1168. doi: 10.1145/2884781.2884877.

[154] V. Saini, H. Sajnani, J. Kim, and C. Lopes, "SourcererCC and SourcererCC-I: tools to detect clones in batch mode and during software development," in *Proceedings of the 38th International Conference on Software Engineering Companion*, New York, NY, USA: ACM, May 2016, pp. 597–600. doi: 10.1145/2889160.2889165.

[155] Y.-L. Hung and S. Takada, "CPPCD: a token-based approach to detecting potential clones," in *2020 IEEE 14th International Workshop on Software Clones (IWSC)*, IEEE, Feb. 2020, pp. 26–32. doi: 10.1109/IWSC50991.2020.9047636.

[156] Y. Wu *et al.*, "SCDetector: software functional clone detection based on semantic tokens analysis," in *Proceedings of the 35th IEEE/ACM International Conference on Automated Software Engineering*, New York, NY, USA: ACM, Dec. 2020, pp. 821–833. doi: 10.1145/3324884.3416562.

[157] P. Wang, J. Svajlenko, Y. Wu, Y. Xu, and C. K. Roy, "CCAligner: a token based large-gap clone detector," in *Proceedings of the 40th International Conference on Software Engineering*, New York, NY, USA: ACM, May 2018, pp. 1066–1077. doi: 10.1145/3180155.3180179.

[158] E. Juergens, F. Deissenboeck, and B. Hummel, "CloneDetective - a workbench for clone detection research," in *2009 IEEE 31st International Conference on Software Engineering*, IEEE, 2009, pp. 603–606. doi: 10.1109/ICSE.2009.5070566.

[159] A. Bhattacharjee and H. M. Jamil, "CodeBlast: A two-stage algorithm for improved program similarity matching in large software repositories," *Proceedings of the ACM Symposium on Applied Computing*, pp. 846–852, 2013, doi: 10.1145/2480362.2480525.

[160] Y. Bian, G. Koru, X. Su, and P. Ma, "SPAPE: A semantic-preserving amorphous procedure extraction method for near-miss clones," *Journal of Systems and Software*, vol. 86, no. 8, pp. 2077–2093, 2013, doi: 10.1016/j.jss.2013.03.061.

[161] Y. Zou, B. Ban, Y. Xue, and Y. Xu, "CCGraph: a PDG-based code clone detector with approximate graph matching," in *2020 35th IEEE/ACM International Conference on Automated Software Engineering (ASE)*, 2020, pp. 931–942.

[162] N. Mehrotra, N. Agarwal, P. Gupta, S. Anand, D. Lo, and R. Purandare, "Modeling functional similarity in source code with graph-based Siamese networks," *IEEE Transactions on Software Engineering*, vol. 48, no. 10, pp. 3771–3789, Oct. 2022, doi: 10.1109/TSE.2021.3105556.

[163] Z. Xue, Z. Jiang, C. Huang, R. Xu, X. Huang, and L. Hu, "SEED: semantic graph based deep detection for type-4 clone," 2022, pp. 120–137. doi: 10.1007/978-3-031-08129-3_8.

[164] Y. Higo, U. Yasushi, M. Nishino, and S. Kusumoto, "Incremental code clone detection: A PDG-based approach," *Proceedings - Working Conference on Reverse Engineering, WCRE*, pp. 3–12, 2011, doi: 10.1109/WCRE.2011.11.

[165] H. A. Basit and S. Jarzabek, "A data mining approach for detecting higher-level clones in software," *IEEE Transactions on Software Engineering*, vol. 35, no. 4, pp. 497–514, Jul. 2009, doi: 10.1109/TSE.2009.16.

[166] M. White, M. Tufano, C. Vendome, and D. Poshyvanyk, "Deep learning code fragments for code clone detection," *ASE 2016 - Proceedings of the 31st IEEE/ACM International Conference on Automated Software Engineering*, pp. 87–98, 2016, doi: 10.1145/2970276.2970326.

[167] M. Hammad, Ö. Babur, H. Abdul Basit, and M. van den Brand, "DeepClone: modeling clones to generate code predictions," 2020, pp. 135–151. doi: 10.1007/978-3-030-64694-3_9.

[168] Y. Zhang and T. Wang, "CCEyes: an effective tool for code clone detection on large-scale open source repositories," in *2021 IEEE International Conference on Information Communication and Software Engineering (ICICSE)*, IEEE, Mar. 2021, pp. 61–70. doi: 10.1109/ICICSE52190.2021.9404141.

[169] A. Zakari, S. P. Lee, K. A. Alam, and R. Ahmad, "Software fault localisation: a systematic mapping study," *IET Software*, vol. 13, no. 1, pp. 60–74, Feb. 2019, doi: 10.1049/iet-sen.2018.5137.

[170] W. E. Wong, R. Gao, Y. Li, R. Abreu, and F. Wotawa, "A survey on software fault localization," *IEEE Transactions on Software Engineering*, vol. 42, no. 8, pp. 707–740, Aug. 2016, doi: 10.1109/TSE.2016.2521368.

[171] R. Just, D. Jalali, and M. D. Ernst, "Defects4J: a database of existing faults to enable controlled testing studies for Java programs," in *Proceedings of the 2014 International Symposium on Software Testing and Analysis - ISSTA 2014*, New York, New York, USA: ACM Press, 2014, pp. 437–440. doi: 10.1145/2610384.2628055.

[172] R. Ferenc, Z. Tóth, G. Ladányi, I. Siket, and T. Gyimóthy, "A public unified bug dataset for java and its assessment regarding metrics and bug prediction," *Software Quality Journal*, vol. 28, no. 4, pp. 1447–1506, 2020, doi: 10.1007/s11219-020-09515-0.

[173] L. Gazzola, D. Micucci, and L. Mariani, "Automatic software repair: a survey," *IEEE Transactions on Software Engineering*, no. June, pp. 1–1, 2017, doi: 10.1109/TSE.2017.2755013.

[174] M. Zakeri-Nasrabadi, S. Parsa, E. Esmaili, and F. Palomba, "A systematic literature review on the code smells datasets and validation mechanisms," *ACM Comput Surv*, May 2023, doi: 10.1145/3596908.

[175] S. Duncan, A. Walker, C. DeHaan, S. Alvord, T. Cerny, and P. Tisnovsky, "Pyclone: a Python code clone test bank generator," 2021, pp. 235–243. doi: 10.1007/978-981-33-6385-4_22.

[176] Microsoft Corporation, "GitHub." https://github.com/ (accessed Jan. 07, 2023).

[177] A. Sheneamer and J. Kalita, "Semantic clone detection using machine learning," in *2016 15th IEEE International Conference on Machine Learning and Applications (ICMLA)*, IEEE, Dec. 2016, pp. 1024–1028. doi: 10.1109/ICMLA.2016.0185.

[178] NetworkX, "NetworkX." https://networkx.github.io/ (accessed Apr. 26, 2019).

[179] F. Pedregosa *et al.*, "Scikit-learn: machine learning in python," *Journal of Machine Learning Research*, vol. 12, pp. 2825–2830, 2011, [Online]. Available: http://jmlr.org/papers/v12/pedregosa11a.html

[180] SciTools, "Understand," 2020. https://www.scitools.com/ (accessed Sep. 11, 2020).

[181] R. FERENC, P. Siket, and M. Schneider, "OpenStaticAnalyzer," *University of Szeged*, 2018. https://github.com/sed-inf-u-szeged/OpenStaticAnalyzer (accessed Jun. 23, 2021).

[182] M. Allamanis, E. T. Barr, C. Bird, and C. Sutton, "Suggesting accurate method and class names," *2015 10th Joint Meeting of the European Software Engineering Conference and the ACM SIGSOFT Symposium on the Foundations of Software Engineering, ESEC/FSE 2015 - Proceedings*, pp. 38–49, 2015, doi: 10.1145/2786805.2786849.

[183] M. Dilhara, A. Ketkar, and D. Dig, "Understanding software-2.0," *ACM Transactions on Software Engineering and Methodology*, vol. 30, no. 4, pp. 1–42, Jul. 2021, doi: 10.1145/3453478.

[184] D. Landman, A. Serebrenik, and J. J. Vinju, "Challenges for static analysis of Java reflection - literature review and empirical study," in *2017 IEEE/ACM 39th International Conference on Software Engineering (ICSE)*, IEEE, May 2017, pp. 507–518. doi: 10.1109/ICSE.2017.53.

[185]   Mayrand, Leblanc, and Merlo, "Experiment on the automatic detection of function clones in a software system using metrics," in *Proceedings of International Conference on Software Maintenance ICSM-96*, IEEE, 1996, pp. 244–253. doi: 10.1109/ICSM.1996.565012.

[186]   R. Komondoor and S. Horwitz, "Using slicing to identify duplication in source code," 2001, pp. 40–56. doi: 10.1007/3-540-47764-0_3.

[187]   Z. Li, S. Lu, S. Myagmar, and Y. Zhou, "CP-Miner: finding copy-paste and related bugs in large-scale software code," *IEEE Transactions on Software Engineering*, vol. 32, no. 3, pp. 176–192, Mar. 2006, doi: 10.1109/TSE.2006.28.

[188]   T. Sager, A. Bernstein, M. Pinzger, and C. Kiefer, "Detecting similar Java classes using tree algorithms," *Proceedings - International Conference on Software Engineering*, pp. 65–71, 2006, doi: 10.1145/1137983.1138000.

[189]   M. Gabel, L. Jiang, and Z. Su, "Scalable detection of semantic clones," *Proceedings - International Conference on Software Engineering*, pp. 321–330, 2008, doi: 10.1145/1368088.1368132.

[190]   R. Falke, P. Frenzel, and R. Koschke, "Empirical evaluation of clone detection using syntax suffix trees," *Empir Softw Eng*, vol. 13, no. 6, pp. 601–643, Dec. 2008, doi: 10.1007/s10664-008-9073-9.

[191]   C. K. Roy, "Detection and analysis of near-miss software clones," in *2009 IEEE International Conference on Software Maintenance*, IEEE, Sep. 2009, pp. 447–450. doi: 10.1109/ICSM.2009.5306301.

[192]   W. S. Evans, C. W. Fraser, and F. Ma, "Clone detection via structural abstraction," *Software Quality Journal*, vol. 17, pp. 309–330, 2009.

[193]   G. M. K. Selim, K. C. Foo, and Y. Zou, "Enhancing source-based clone detection using intermediate representation," in *2010 17th Working Conference on Reverse Engineering*, IEEE, Oct. 2010, pp. 227–236. doi: 10.1109/WCRE.2010.33.

[194]   S. U. Rehman, K. Khan, S. Fong, and R. Biuk-Aghai, "An efficient new multi-language clone detection approach from large source code," *Conf Proc IEEE Int Conf Syst Man Cybern*, pp. 937–940, 2012, doi: 10.1109/ICSMC.2012.6377848.

[195]   R. Tekchandani, R. K. Bhatia, and M. Singh, "Semantic code clone detection using parse trees and grammar recovery," *IET Conference Publications*, vol. 2013, no. 647 CP, pp. 41–46, 2013, doi: 10.1049/cp.2013.2291.

[196]   R. K. Tekchandani and K. Raheja, "An efficient code clone detection model on Java byte code using hybrid approach," in *Confluence 2013: The Next Generation Information Technology Summit (4th International Conference)*, Institution of Engineering and Technology, 2013, pp. 1.04-1.04. doi: 10.1049/cp.2013.2287.

[197]   A. Agrawal and S. K. Yadav, "A hybrid-token and textual based approach to find similar code segments," *2013 4th International Conference on Computing, Communications and Networking Technologies, ICCCNT 2013*, pp. 4–7, 2013, doi: 10.1109/ICCCNT.2013.6726700.

[198]   H. Murakami, K. Hotta, Y. Higo, H. Igaki, and S. Kusumoto, "Gapped code clone detection with lightweight source code analysis," in *2013 21st International Conference on Program Comprehension (ICPC)*, IEEE, May 2013, pp. 93–102. doi: 10.1109/ICPC.2013.6613837.

[199]   R. K. Saha, C. K. Roy, and K. A. Schneider, "gCad: A near-miss clone genealogy extractor to support clone evolution analysis," in *2013 IEEE International Conference on Software Maintenance*, IEEE, Sep. 2013, pp. 488–491. doi: 10.1109/ICSM.2013.79.

[200]   S. Cesare, Y. Xiang, and J. Zhang, "Clonewise – detecting package-level clones using machine learning," 2013, pp. 197–215. doi: 10.1007/978-3-319-04283-1_13.

[201]   Y. Higo and S. Kusumoto, "How should we measure functional sameness from Program source code? An exploratory study on Java methods," *Proceedings of the ACM SIGSOFT Symposium on the Foundations of Software Engineering*, vol. 16-21-Nove, pp. 294–305, 2014, doi: 10.1145/2635868.2635886.

[202]   E. Kodhai and S. Kanmani, "Method-level code clone detection through LWH (light weight hybrid) approach," *Journal of Software Engineering Research and Development*, vol. 2, no. 1, pp. 1–29, 2014, doi: 10.1186/s40411-014-0012-8.

[203]   A. Avetisyan, S. Kurmangaleev, S. Sargsyan, M. Arutunian, and A. Belevantsev, "LLVM-based code clone detection framework," in *2015 Computer Science and Information Technologies (CSIT)*, IEEE, Sep. 2015, pp. 100–104. doi: 10.1109/CSITechnol.2015.7358250.

[204]	I. Keivanloo, F. Zhang, and Y. Zou, "Threshold-free code clone detection for a large-scale heterogeneous Java repository," in *2015 IEEE 22nd International Conference on Software Analysis, Evolution, and Reengineering (SANER)*, IEEE, Mar. 2015, pp. 201–210. doi: 10.1109/SANER.2015.7081830.

[205]	J. Kim, T. G. Kim, and E. G. Im, "Structural information based malicious app similarity calculation and clustering," in *Proceedings of the 2015 Conference on research in adaptive and convergent systems*, New York, NY, USA: ACM, Oct. 2015, pp. 314–318. doi: 10.1145/2811411.2811545.

[206]	T. Kamiya, "An execution-semantic and content-and-context-based code-clone detection and analysis," *2015 IEEE 9th International Workshop on Software Clones, IWSC 2015 - Proceedings*, pp. 1–7, 2015, doi: 10.1109/IWSC.2015.7069882.

[207]	B. Joshi, P. Budhathoki, W. L. Woon, and D. Svetinovic, "Software clone detection using clustering approach," 2015, pp. 520–527. doi: 10.1007/978-3-319-26535-3_59.

[208]	T. Schmorleiz and R. Lämmel, "Similarity management of 'cloned and owned' variants," in *Proceedings of the 31st Annual ACM Symposium on Applied Computing*, New York, NY, USA: ACM, Apr. 2016, pp. 1466–1471. doi: 10.1145/2851613.2851785.

[209]	M. Sudhamani and L. Rangarajan, "Code clone detection based on order and content of control statements," *Proceedings of the 2016 2nd International Conference on Contemporary Computing and Informatics, IC3I 2016*, pp. 59–64, 2016, doi: 10.1109/IC3I.2016.7917935.

[210]	A. Chandran, L. Jain, S. Rawat, and K. Srinathan, "Discovering vulnerable functions: a code similarity based approach," 2016, pp. 390–402. doi: 10.1007/978-981-10-2738-3_34.

[211]	L. Li, H. Feng, W. Zhuang, N. Meng, and B. Ryder, "CCLearner: a deep learning-based clone detection approach," in *2017 IEEE International Conference on Software Maintenance and Evolution (ICSME)*, IEEE, Sep. 2017, pp. 249–260. doi: 10.1109/ICSME.2017.46.

[212]	C. V Lopes *et al.*, "DéjàVu: a map of code duplicates on GitHub," *Proceedings of the ACM on Programming Languages*, vol. 1, no. OOPSLA, pp. 1–28, 2017.

[213]	T. Lavoie, M. Mérineau, E. Merlo, and P. Potvin, "A case study of TTCN-3 test scripts clone analysis in an industrial telecommunication setting," *Inf Softw Technol*, vol. 87, pp. 32–45, 2017, doi: 10.1016/j.infsof.2017.01.008.

[214]	C. Ragkhitwetsagul and J. Krinke, "Using compilation/decompilation to enhance clone detection," in *2017 IEEE 11th International Workshop on Software Clones (IWSC)*, IEEE, Feb. 2017, pp. 1–7. doi: 10.1109/IWSC.2017.7880502.

[215]	D. Zou *et al.*, "SCVD: a new semantics-based approach for cloned vulnerable code detection," 2017, pp. 325–344. doi: 10.1007/978-3-319-60876-1_15.

[216]	M. Tufano, C. Watson, G. Bavota, M. di Penta, M. White, and D. Poshyvanyk, "Deep learning similarities from different representations of source code," in *2018 IEEE/ACM 15th International Conference on Mining Software Repositories (MSR)*, 2018, pp. 542–553.

[217]	G. Mostaeen, J. Svajlenko, B. Roy, C. K. Roy, and K. A. Schneider, "On the use of machine learning techniques towards the design of cloud based automatic code clone validation tools," *Proceedings - 18th IEEE International Working Conference on Source Code Analysis and Manipulation, SCAM 2018*, pp. 155–164, 2018, doi: 10.1109/SCAM.2018.00025.

[218]	R. Tajima, M. Nagura, and S. Takada, "Detecting functionally similar code within the same project," in *2018 IEEE 12th International Workshop on Software Clones (IWSC)*, IEEE, Mar. 2018, pp. 51–57. doi: 10.1109/IWSC.2018.8327319.

[219]	T. Vislavski, G. Rakic, N. Cardozo, and Z. Budimac, "LICCA: A tool for cross-language clone detection," in *2018 IEEE 25th International Conference on Software Analysis, Evolution and Reengineering (SANER)*, IEEE, Mar. 2018, pp. 512–516. doi: 10.1109/SANER.2018.8330250.

[220]	Y. Wang and D. Liu, "Image-based clone code detection and visualization," in *2019 International Conference on Artificial Intelligence and Advanced Manufacturing (AIAM)*, IEEE, Oct. 2019, pp. 168–175. doi: 10.1109/AIAM48774.2019.00041.

[221]	Y. Gao, Z. Wang, S. Liu, L. Yang, W. Sang, and Y. Cai, "TECCD: A Tree Embedding Approach for Code Clone Detection," *Proceedings - 2019 IEEE International Conference on Software Maintenance and Evolution, ICSME 2019*, pp. 145–156, 2019, doi: 10.1109/ICSME.2019.00025.

[222]    J. Zeng, K. Ben, X. Li, and X. Zhang, "Fast code clone detection based on weighted recursive autoencoders," *IEEE Access*, vol. 7, pp. 125062–125078, 2019, doi: 10.1109/ACCESS.2019.2938825.

[223]    J. Zhang, X. Wang, H. Zhang, H. Sun, K. Wang, and X. Liu, "A novel neural source code representation based on abstract syntax tree," in *2019 IEEE/ACM 41st International Conference on Software Engineering (ICSE)*, IEEE, May 2019, pp. 783–794. doi: 10.1109/ICSE.2019.00086.

[224]    G. Mostaeen, J. Svajlenko, B. Roy, C. K. Roy, and K. A. Schneider, "CloneCognition: machine learning based code clone validation tool," in *Proceedings of the 2019 27th ACM Joint Meeting on European Software Engineering Conference and Symposium on the Foundations of Software Engineering*, New York, NY, USA: ACM, Aug. 2019, pp. 1105–1109. doi: 10.1145/3338906.3341182.

[225]    M. Sudhamani and L. Rangarajan, "Code similarity detection through control statement and program features," *Expert Syst Appl*, vol. 132, pp. 63–75, 2019, doi: 10.1016/j.eswa.2019.04.045.

[226]    D. Tukaram and B. Uma Maheswari, "Design and development of software tool for code clone search, detection, and analysis," *Proceedings of the 3rd International Conference on Electronics and Communication and Aerospace Technology, ICECA 2019*, pp. 1002–1006, 2019, doi: 10.1109/ICECA.2019.8821928.

[227]    J. Yang, Y. Xiong, and J. Ma, "A function level Java code clone detection method," *Proceedings of 2019 IEEE 4th Advanced Information Technology, Electronic and Automation Control Conference, IAEAC 2019*, no. Iaeac, pp. 2128–2134, 2019, doi: 10.1109/IAEAC47372.2019.8998079.

[228]    M. Gharehyazie, B. Ray, M. Keshani, M. S. Zavosht, A. Heydarnoori, and V. Filkov, "Cross-project code clones in GitHub," *Empir Softw Eng*, vol. 24, no. 3, pp. 1538–1573, Jun. 2019, doi: 10.1007/s10664-018-9648-z.

[229]    G. Li *et al.*, "SAGA: efficient and large-scale detection of near-miss clones with GPU acceleration," in *2020 IEEE 27th International Conference on Software Analysis, Evolution and Reengineering (SANER)*, IEEE, Feb. 2020, pp. 272–283. doi: 10.1109/SANER48275.2020.9054832.

[230]    Y. Yuan, W. Kong, G. Hou, Y. Hu, M. Watanabe, and A. Fukuda, "From local to global semantic clone detection," in *2019 6th International Conference on Dependable Systems and Their Applications (DSA)*, IEEE, Jan. 2020, pp. 13–24. doi: 10.1109/DSA.2019.00012.

[231]    P. M. Caldeira, K. Sakamoto, H. Washizaki, Y. Fukazawa, and T. Shimada, "Improving syntactical clone detection methods through the use of an intermediate representation," in *2020 IEEE 14th International Workshop on Software Clones (IWSC)*, IEEE, Feb. 2020, pp. 8–14. doi: 10.1109/IWSC50091.2020.9047637.

[232]    H. Xue, Y. Mei, K. Gogineni, G. Venkataramani, and T. Lan, "Twin-Finder: integrated reasoning engine for pointer-related code clone detection," in *2020 IEEE 14th International Workshop on Software Clones (IWSC)*, IEEE, Feb. 2020, pp. 1–7. doi: 10.1109/IWSC50091.2020.9047638.

[233]    M. Wu, P. Wang, K. Yin, H. Cheng, Y. Xu, and C. K. Roy, "LVMapper: a large-variance cone detector using sequencing alignment approach," *IEEE Access*, vol. 8, pp. 27986–27997, 2020, doi: 10.1109/ACCESS.2020.2971545.

[234]    G. Mostaeen, B. Roy, C. K. Roy, K. Schneider, and J. Svajlenko, "A machine learning based framework for code clone validation," *Journal of Systems and Software*, vol. 169, p. 110686, 2020, doi: 10.1016/j.jss.2020.110686.

[235]    W. Dong, Z. Feng, H. Wei, and H. Luo, "A novel code stylometry-based code clone detection strategy," in *2020 International Wireless Communications and Mobile Computing (IWCMC)*, IEEE, Jun. 2020, pp. 1516–1521. doi: 10.1109/IWCMC48107.2020.9148302.

[236]    A. Zhang, K. Liu, L. Fang, Q. Liu, X. Yun, and S. Ji, "Learn to align: a code alignment network for code clone detection," in *2021 28th Asia-Pacific Software Engineering Conference (APSEC)*, IEEE, Dec. 2021, pp. 1–11. doi: 10.1109/APSEC53868.2021.00008.

[237]    W. Amme, T. S. Heinze, and A. Schafer, "You look so different: finding structural clones and subclones in Java source code," in *2021 IEEE International Conference on Software Maintenance and Evolution (ICSME)*, IEEE, Sep. 2021, pp. 70–80. doi: 10.1109/ICSME52107.2021.00013.

[238]    N. D. Q. Bui, Y. Yu, and L. Jiang, "InferCode: self-supervised learning of code representations by predicting subtrees," in *2021 IEEE/ACM 43rd International Conference on Software Engineering (ICSE)*, IEEE, May 2021, pp. 1186–1197. doi: 10.1109/ICSE43902.2021.00109.

[239] W. Hua, Y. Sui, Y. Wan, G. Liu, and G. Xu, "FCCA: hybrid code representation for functional clone detection using attention networks," *IEEE Trans Reliab*, vol. 70, no. 1, pp. 304–318, Mar. 2021, doi: 10.1109/TR.2020.3001918.

[240] A. Schafer, W. Amme, and T. S. Heinze, "Stubber: compiling source code into bytecode without dependencies for Java code clone detection," in *2021 IEEE 15th International Workshop on Software Clones (IWSC)*, IEEE, Oct. 2021, pp. 29–35. doi: 10.1109/IWSC53727.2021.00011.

[241] A. Sheneamer, S. Roy, and J. Kalita, "An effective semantic code clone detection framework using pairwise feature fusion," *IEEE Access*, vol. 9, pp. 84828–84844, 2021, doi: 10.1109/ACCESS.2021.3079156.

[242] S. B. Ankali and L. Parthiban, "Development of porting analyzer to search cross-language code clones using levenshtein distance," 2021, pp. 623–632. doi: 10.1007/978-981-16-0878-0_60.

[243] H. Jin, Z. Cui, S. Liu, and L. Zheng, "Improving code clone detection accuracy and efficiency based on code complexity analysis," in *2022 9th International Conference on Dependable Systems and Their Applications (DSA)*, IEEE, Aug. 2022, pp. 64–72. doi: 10.1109/DSA56465.2022.00017.

[244] C. Tao, Q. Zhan, X. Hu, and X. Xia, "C4: contrastive cross-language code clone detection," 2022.

[245] Z. Li *et al.*, "Unleashing the power of compiler intermediate representation to enhance neural program embeddings," in *Proceedings of the 44th International Conference on Software Engineering*, New York, NY, USA: ACM, May 2022, pp. 2253–2265. doi: 10.1145/3510003.3510217.

[246] M. Hammad, O. Babur, H. A. Basit, and M. van den Brand, "Clone-seeker: effective code clone search using annotations," *IEEE Access*, vol. 10, pp. 11696–11713, 2022, doi: 10.1109/ACCESS.2022.3145686.

[247] S. Karthik and B. Rajdeepa, "A collaborative method for code clone detection using a deep learning model," *Advances in Engineering Software*, vol. 174, p. 103327, Dec. 2022, doi: 10.1016/j.advengsoft.2022.103327.

[248] M. Chochlov *et al.*, "Using a nearest-neighbour, BERT-based approach for scalable clone detection," in *2022 IEEE International Conference on Software Maintenance and Evolution (ICSME)*, IEEE, Oct. 2022, pp. 582–591. doi: 10.1109/ICSME55016.2022.00080.

[249] F. Leone and S. Takada, "Towards overcoming type limitations in semantic clone detection," in *2022 IEEE 16th International Workshop on Software Clones (IWSC)*, IEEE, Oct. 2022, pp. 25–31. doi: 10.1109/IWSC55060.2022.00013.

[250] M. H. Islam, R. Paul, and M. Mondal, "Predicting buggy code clones through machine learning," in *Proceedings of the 32nd Annual International Conference on Computer Science and Software Engineering*, 2022, pp. 130–139. doi: https://dl.acm.org/doi/10.5555/3566055.3566070.

[251] Y. Hu, D. Zou, J. Peng, Y. Wu, J. Shan, and H. Jin, "TreeCen: Building tree graph for scalable semantic code clone detection," in *37th IEEE/ACM International Conference on Automated Software Engineering*, New York, NY, USA: ACM, Oct. 2022, pp. 1–12. doi: 10.1145/3551349.3556927.

[252] Y. Wu, S. Feng, D. Zou, and H. Jin, "Detecting semantic code clones by building AST-based Markov chains model," in *37th IEEE/ACM International Conference on Automated Software Engineering*, New York, NY, USA: ACM, Oct. 2022, pp. 1–13. doi: 10.1145/3551349.3560426.

[253] H. Thaller, L. Linsbauer, and A. Egyed, "Semantic clone detection via probabilistic software modeling," 2022, pp. 288–309. doi: 10.1007/978-3-030-99429-7_16.

[254] L. Yang *et al.*, "FastDCF: A partial index based distributed and scalable near-miss code clone detection approach for very large code repositories," 2022, pp. 210–222. doi: 10.1007/978-3-030-96772-7_20.

[255] S. Patel and R. Sinha, "Combining holistic source code representation with siamese neural networks for detecting code clones," 2022, pp. 148–159. doi: 10.1007/978-3-031-04673-5_12.

[256] X. Guo, R. Zhang, L. Zhou, and X. Lu, "Precise code clone detection with architecture of abstract syntax trees," 2022, pp. 117–126. doi: 10.1007/978-3-031-19211-1_10.

[257] Y. Li, C. Yu, and Y. Cui, "TPCaps: a framework for code clone detection and localization based on improved CapsNet," *Applied Intelligence*, Dec. 2022, doi: 10.1007/s10489-022-03158-3.

[258] A. Zhang, L. Fang, C. Ge, P. Li, and Z. Liu, "Efficient transformer with code token learner for code clone detection," *Journal of Systems and Software*, vol. 197, p. 111557, Mar. 2023, doi: 10.1016/j.jss.2022.111557.

[259]  O. Ehsan, F. Khomh, Y. Zou, and D. Qiu, "Ranking code clones to support maintenance activities," *Empir Softw Eng*, vol. 28, no. 3, p. 70, Jun. 2023, doi: 10.1007/s10664-023-10292-0.

[260]  W. Wang, Z. Deng, Y. Xue, and Y. Xu, "CCStokener: fast yet accurate code clone detection with semantic token," *Journal of Systems and Software*, vol. 199, p. 111618, May 2023, doi: 10.1016/j.jss.2023.111618.